\documentclass[11pt,onecolumn,a4paper,unpublished]{quantumarticle}
\pdfoutput=1 

\usepackage{tikz}
\usetikzlibrary{positioning, calc}
\usetikzlibrary{arrows}
\usepackage{graphicx}
\usepackage{mdwlist}
\usepackage{thmtools}
\usepackage{amssymb}
\usepackage{physics}
\usepackage{amsmath}
\usepackage{array}
\usepackage{url}
\usepackage[numbers]{natbib}
\usetikzlibrary{fadings}
\usetikzlibrary{decorations.pathmorphing}
\usepackage{authblk}

\usetikzlibrary{decorations.pathreplacing}
\tikzset{snake it/.style={decorate, decoration=snake}}

\usetikzlibrary{decorations.pathreplacing,decorations.markings}

\usepackage[dvipsnames]{xcolor}
\usepackage[colorlinks=true,
            linkcolor=blue,
            citecolor=blue,
            urlcolor=blue]{hyperref}

\usepackage{float}
\usepackage{subcaption}

\usepackage{cleveref}

\usepackage[T1]{fontenc}

\newcommand{\A}{\mathcal{A}}
\newcommand{\N}{\mathcal{N}}
\newcommand{\U}{\mathcal{U}}

\DeclareMathOperator{\supp}{supp}

\tikzset{
    >=stealth',
    punkt/.style={
           rectangle,
           rounded corners,
           draw=black, very thick,
           text width=6.5em,
           minimum height=2em,
           text centered},
    pil/.style={
           ->,
           thick,
           shorten <=2pt,
           shorten >=2pt,},
  on each segment/.style={
    decorate,
    decoration={
      show path construction,
      moveto code={},
      lineto code={
        \path [#1]
        (\tikzinputsegmentfirst) -- (\tikzinputsegmentlast);
      },
      curveto code={
        \path [#1] (\tikzinputsegmentfirst)
        .. controls
        (\tikzinputsegmentsupporta) and (\tikzinputsegmentsupportb)
        ..
        (\tikzinputsegmentlast);
      },
      closepath code={
        \path [#1]
        (\tikzinputsegmentfirst) -- (\tikzinputsegmentlast);
      },
    },
  },
  mid arrow/.style={postaction={decorate,decoration={
        markings,
        mark=at position .5 with {\arrow[#1]{stealth'}}
      }}}
}

\newtheorem{theorem}{Theorem}
\newtheorem{conjecture}[theorem]{Conjecture}
\newtheorem{definition}[theorem]{Definition}
\newtheorem{lemma}[theorem]{Lemma}
\newtheorem{remark}[theorem]{Remark}
\newenvironment{proof}[1][Proof]{\noindent\textbf{#1.}}{\ \rule{0.5em}{0.5em}}

\begin{document} 

\title{Entanglement sharing schemes}

\newcommand{\PI}{Perimeter Institute for Theoretical Physics, Waterloo, Ontario, Canada}
\newcommand{\IQC}{Institute for Quantum Computing, University of Waterloo, Ontario, Canada}
\newcommand{\NRC}{National Research Council Canada, Waterloo, Ontario, Canada}
\newcommand{\Rikkyo}{Department of Physics, Rikkyo University, Tokyo, Japan}
\newcommand{\Freie}{Dahlem Center for Complex Quantum Systems, Freie Universit\"{a}t Berlin, Berlin, Germany}
\newcommand{\UW}{Department of Physics and Astronomy, University of Waterloo, Ontario, Canada}

\author{Zahra Khanian}
\affiliation{\PI}
\orcid{}

\author{Dongjin Lee}
\affiliation{\PI}
\affiliation{\UW}
\orcid{}

\author{Debbie Leung}
\orcid{0000-0003-3750-2648}
\affiliation{\IQC}
\affiliation{\PI}

\author{Zhi Li}
\thanks{current affiliation: IBM Quantum}
\email{zli@ibm.com}
\affiliation{\NRC}
\affiliation{\PI}

\author{Alex May}
\email{amay@perimeterinstitute.ca}
\orcid{0000-0002-4030-5410}
\affiliation{\PI}
\affiliation{\IQC}

\author{Stanley Miao}
\orcid{0009-0000-7930-7563}
\affiliation{\PI}
\affiliation{\IQC}

\author{Takato Mori}
\orcid{}
\affiliation{\PI}
\affiliation{\Rikkyo}

\author{Farzin Salek}
\orcid{}
\affiliation{\IQC}
\affiliation{\Freie}

\author{Jinmin Yi}
\orcid{0000-0001-6948-2464}
\affiliation{\PI}
\affiliation{\UW}

\author{Beni Yoshida}
\orcid{}
\affiliation{\PI}

\abstract{
We ask how quantum correlations can be distributed among many subsystems. 
To address this, we define entanglement sharing schemes (ESS) where certain pairs of subsystems allow entanglement to be recovered via local operations, while other pairs must not. 
Entanglement sharing schemes come in two variants, one where the partner system with which entanglement should be prepared is known, and one where it is not. 
In the case of known partners, we fully characterize the access structures realizable for ESS when using stabilizer states, construct efficient schemes for threshold access structures, and give a conjecture for the access structures realizable with general states. 
In the unknown partner case, we again give a complete characterization in the stabilizer setting, additionally give a complete characterization of the case where there are no restrictions on unauthorized pairs, and we prove a set of necessary conditions on general schemes which we conjecture are also sufficient. 
Finally, we give an application of the theory of entanglement sharing to resolve an open problem related to the distribution of entanglement in response to time-sensitive requests in quantum networks. 
}

\vfill

\maketitle

\pagebreak

\tableofcontents

\section{Introduction}

In this work, we introduce and study a new framework, which we call \emph{entanglement sharing schemes}\footnote{The term `entanglement sharing scheme' was also used in \cite{choi2013entanglement}, where it refers to a related setting. Since what were previously called entanglement sharing schemes form a subset of the schemes we consider, we believe the reuse of the term is justified.}.
Entanglement sharing schemes provide flexible means of storing, protecting, and regulating access to entanglement among multiple parties. 
We illustrate one application of our constructions to an open problem in entanglement summoning \cite{adlam2018relativistic,dolev2021distributing}, concerning how entanglement can be made available in response to time-sensitive requests in quantum networks.
From a more fundamental perspective, we view ESS as capturing the essential principles governing how entanglement is stored and distributed across multipartite quantum states.

To set the stage for our work, it is helpful to recall the framework of quantum secret sharing (QSS) \cite{cleve1999share,gottesman2000theory}, which serves as our primary inspiration.
Secret sharing is a fundamental tool in both classical and quantum cryptography. 
The setting is as follows.
A designated dealer is given an unknown quantum state $\ket{\psi}$ encoded into system $A$. 
The dealer then applies an encoding map that distributes $A$ into shares $S_1,\ldots,S_n$, where share $S_i$ is sent to party $i$.
The encoding is required to respect an \emph{access structure}, specified by a set $\mathcal{A}$ of \emph{authorized sets} of parties and a set $\mathcal{U}$ of \emph{unauthorized sets}.
Any subset $A_i\in\mathcal{A}$ should, by combining their shares, be able to reconstruct the state $\ket{\psi}$, while any subset $U_i\in\mathcal{U}$ should be unable to learn anything about $\ket{\psi}$.

Beyond its cryptographic applications, quantum secret sharing can also be viewed more fundamentally as a framework for describing how quantum information is distributed across multiple subsystems.
It addresses the question: under what conditions can the same quantum information be stored within authorized collections of subsystems $A_1,\ldots,A_n\in\mathcal{A}$ while remaining inaccessible to unauthorized collections $U_1,\ldots,U_m\in\mathcal{U}$?\footnote{This perspective is emphasized, for instance, in \cite{hayden2016summoning,hayden2019localizing}.}

Building on this perspective, we ask how \emph{quantum correlations} can be distributed among many subsystems. 
To formalize this idea, we introduce the notion of a \emph{pair access structure}.
Such a structure consists of a set of authorized pairs $\mathcal{A}$ and a set of unauthorized pairs $\mathcal{U}$. 
Each pair consists of two disjoint subsets $T_i,T_j$ of the full set of parties $\Gamma$.
For authorized pairs $\{T_i,T_j\}\in\mathcal{A}$, there should exist local operations that recover a system $A$ from $T_i$ and a system $B$ from $T_j$, such that $AB$ is prepared in the desired quantum state $|\psi\rangle_{AB}$.
For unauthorized pairs, no such process should yield output systems $C$ and $D$ and a state that approximates $|\psi\rangle_{CD}$ to high fidelity.\footnote{More precisely, we give two notions of security in the main text: \emph{weak} and \emph{strong} security. In weak security, we require $\ket{\psi}$ is not recovered perfectly. In strong security, we require no output approximating $\ket{\psi}$ to bounded error can be produced.}

Within this framework, several natural variants arise.
First, the target state $\ket{\psi}$ (which has two subsystems) may either be known to all parties in advance or be an unknown state provided as input to the scheme.
In this work we focus on the case where the state is known, which we refer to as an \emph{entanglement sharing scheme} (ESS).
Note that in quantum secret sharing the known-state case is trivial, since any subsets can always prepare the known state and hence no nontrivial unauthorized subsets can exist.
By contrast, in the setting of correlations the situation is more subtle: the monogamy of entanglement and the impossibility of generating entanglement between non-communicating parties together impose strong restrictions, enabling the existence of unauthorized pairs and permitting a wide variety of nontrivial pair access structures.

Another subtlety in the setting of pair access structures is whether parties know which other set of parties they are expected to share entanglement with.
As a simple example, consider three parties $P_1$, $P_2$, and $P_3$, where an EPR pair is distributed between $P_1$ and $P_2$ and a second EPR pair between $P_1$ and $P_3$.
If the partner is known, we may realize an access structure such that both $\{\{P_1\},\{P_2\}\}\in\mathcal{A}$ and $\{\{P_1\},\{P_3\}\}\in\mathcal{A}$, since once $P_1$ is told which partner to be entangled with, it can simply return the corresponding subsystem.
By contrast, if the partner is unknown, $P_1$ only knows that they must return a system entangled with someone, but not with whom. 
The monogamy of entanglement then prevents $P_1$ from consistently returning a system which is maximally entangled with the other returned system (the system returned by $P_2$ or $P_3$), and in fact this access structure cannot be realized in the unknown-partner setting.

Another possible variation of our setting would be to allow classical communication between the party holding $T_i$ and the party holding $T_j$. 
This scenario seems well defined and interesting in the known partner setting, but we leave studying it to future work. 
For the unknown partner case we could not allow classical communication, since that would reveal the identity of the partner. 

In this work, we study the access structures achievable in entanglement sharing schemes under both the known-partner and unknown-partner settings.
We provide a set of necessary conditions in each case.
Within the class of stabilizer states \cite{gottesman1996}, we fully characterize the realizable access structures, and we present an explicit example showing that non-stabilizer constructions enable additional structures to be realized.
In addition, we construct schemes that realize general access structures under a weakened notion of security, where unauthorized pairs may become nearly authorized in a sense that we make precise.
A complete understanding of the strong security setting remains an open problem.
In particular, we leave for future work a full characterization of the non-stabilizer case under the strong notion of security.

\begin{figure}
\centering
\begin{tikzpicture}[scale=1.6,rotate=19]
\begin{scope}[thick,decoration={
    markings,
    mark=at position 0.5 with {\arrow{triangle 45}}}
    ] 
  
\draw[fill=black] (1,0) circle (0.05cm);
\node[right] at (1,0) {$D_1$};

\draw[fill=black] (0.309,0.951) circle (0.05cm);
\node[above] at (0.309,0.951) {$D_2$};

\draw[fill=black] (-0.809,0.588) circle (0.05cm);
\node[left] at (-0.809,0.588) {$D_3$};

\draw[fill=black] (-0.809,-0.588) circle (0.05cm);
\node[below left] at (-0.809,-0.588) {$D_4$};

\draw[fill=black] (0.309,-0.951) circle (0.05cm);
\node[below right] at (0.309,-0.951) {$D_5$};

\begin{scope} [rotate=0]
\draw (1,0) -- (0.309,0.951);
\draw[-triangle 45] (1,0) -> (0.481,0.713);
\draw[-triangle 45] (0.309,0.951) -> (0.827,0.238);
\end{scope}

\begin{scope} [rotate=72]
\draw (1,0) -- (0.309,0.951);
\draw[-triangle 45] (1,0) -> (0.481,0.713);
\draw[-triangle 45] (0.309,0.951) -> (0.827,0.238);
\end{scope}

\begin{scope} [rotate=72*2]
\draw (1,0) -- (0.309,0.951);
\draw[-triangle 45] (1,0) -> (0.481,0.713);
\draw[-triangle 45] (0.309,0.951) -> (0.827,0.238);
\end{scope}

\begin{scope} [rotate=72*3]
\draw (1,0) -- (0.309,0.951);
\draw[-triangle 45] (1,0) -> (0.481,0.713);
\draw[-triangle 45] (0.309,0.951) -> (0.827,0.238);
\end{scope}

\begin{scope} [rotate=72*4]
\draw (1,0) -- (0.309,0.951);
\draw[-triangle 45] (1,0) -> (0.481,0.713);
\draw[-triangle 45] (0.309,0.951) -> (0.827,0.238);
\end{scope}

\end{scope}
\end{tikzpicture}
\caption{A five-vertex graph. The five vertices represent spatially separated labs. In the entanglement summoning setting, two of the five labs receive a request to return the maximally entangled $\frac{1}{\sqrt{d}}\sum_i \ket{i}\ket{i}$ state. Monogamy of entanglement and the limited communication between neighbours make this challenging: if a request is received at $D_1$ for example, the lab there cannot distinguish between the partner request being at $D_3$ or at $D_4$. Thus they may share entanglement with both labs but do not know which system in their lab is entangled with the requested partner. Constraints on entanglement sharing schemes show completing this task is impossible.}
\label{fig:pentagonintro}
\end{figure}
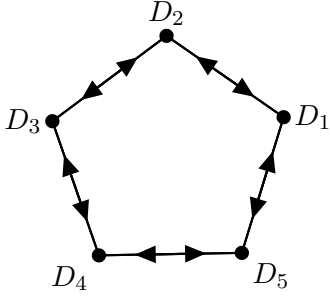

The study of entanglement sharing schemes was first motivated for us by an application to \emph{entanglement summoning} \cite{adlam2018relativistic,dolev2021distributing}. In fact, our results resolve an open problem posed in \cite{dolev2021distributing}.
In entanglement summoning, a collection of spatially distributed labs must respond to time-sensitive requests to return specific quantum states (in our case, maximally entangled states).
The time constraints prevent global communication among all labs, though communication may still be possible between certain nearby labs.
By applying the framework of entanglement sharing schemes, we characterize a previously unresolved instance of entanglement summoning, illustrated in \cref{fig:pentagonintro}.
We show that the constraints imposed by entanglement sharing schemes in the unknown-partner setting rule out the possibility of implementing this task perfectly.

\subsection{First example}\label{sec:firstexample}

To illustrate the quantum states underlying our constructions, consider the following example, in the known partner setting, of an entanglement sharing scheme.
The scheme distributes an entangled state among five parties, each holding a 5-dimensional qudit:
\begin{align}\label{eq:225}
\ket{\Psi}_{ABCDE} = \frac{1}{5}\sum_{k,s} \ket{k}_A \ket{k+s}_B \ket{k+2s}_C \ket{k+3s}_D \ket{k+4s}_E ,
\end{align}
where $k,s\in\mathbb{F}_5$.
This state realizes what we call a $((2,2,5))$ threshold access structure for entanglement sharing with a known partner.
In this structure, there are a total of 5 shares, and every pair $\{T_i,T_j\}$ with $T_i,T_j$ each containing two shares ($|T_i|=|T_j|=2$) is authorized in the sense that they can recover a maximally entangled state via local operations, while all pairs with $|T_i|=1$, $|T_j|\leq 2$ are unauthorized.
The scheme is obtained by taking a $(2,5)$ threshold classical secret sharing scheme and forming a superposition over all choices of secret and randomness.
This approach generalizes to yield efficient (small local dimension) constructions of $((a,b,a+2b-1))$ threshold access structures, where every pair $\{T_i,T_j\}$ with $|T_i|=a$ and $|T_j|=b$ is authorized, while smaller pairs are not.

The construction of the $((2,2,5))$ scheme is surprisingly efficient.
To appreciate this, consider a more naive approach. 
Suppose $\{T_i,T_j\}$ and $\{T_i,T_k\}$ are both authorized pairs.
Then $T_i$ must share entanglement with both $T_j$ and $T_k$.
A straightforward way to achieve this is to distribute two independent EPR pairs—one for $\{T_i,T_j\}$ and another for $\{T_i,T_k\}$.
However, this already forces the dimension of $T_i$ to be at least $d_E^2$, for $d_E$ the dimension of the recovered entangled state.
Continuing in this manner, ensuring entanglement for all authorized pairs $\{T_i, T_j\}$ with $|T_i|=|T_j|=2$ would lead to a prohibitively large state.
In the $((2,2,5))$ example, this would lead to $T_i$ having dimension at least $d_E^3$, since once we choose two shares to make up $T_i$, there are three distinct ways to choose $T_j$ such that $\{T_i,T_j\}$ is authorized.
By contrast, in the threshold structure described above, a more efficient construction is possible: each share has dimension $d_E$, so each $T_i$ has dimension $d_E^2$.
Moreover, we show that the subsystem sizes in this construction are close to optimal for realizing the $((2,2,5))$ access structure, in that the dimension of each system must be at least the square root of the dimension of the shared entangled state.\footnote{This follows by applying the lower bound on significant shares from \cref{sec:lowerbounds}.}

In the unknown partner case, the conditions under which an access structure can be realized change significantly. 
Consider again the example of the $((2,2,5))$ threshold access structure. 
Whereas in the known partner case \cref{eq:225} provides an explicit construction, this access structure cannot be realized in the unknown partner setting. 
To see this suppose, by way of contradiction, that $\{AB,CD\}$ is authorized, which means there are decoding unitaries $U_{AB}$ and $U_{CD}$ acting on $AB$ and $CD$ producing systems $X$ and $Y$ such that $\ket{\psi}_{XY}$ is an EPR pair. 
Then this also means that\footnote{Here we are using the ``transpose trick'', which states that for any $\mathcal{O}$, we have $\mathcal{O}_{X}\ket{\Psi^+}_{XY}=\mathcal{O}_Y^T\ket{\Psi^+}_{XY}$.}
\begin{align}
    \forall \,V,\,\,\, \,\,U_{AB\rightarrow XE}^\dagger V_X U_{AB\rightarrow XE} \ket{\Psi}_{ABCDE} = U_{CD\rightarrow YE'}^\dagger V_Y^T U_{CD\rightarrow YE'} \ket{\Psi}_{ABCDE} 
\end{align}
or more briefly we define $\tilde{V}_{AB}=U_{AB\rightarrow XE}^\dagger V_X U_{AB\rightarrow XE}$, $\tilde{V}^T_{CD}=U_{CD\rightarrow YE'}^\dagger V^T_Y U_{CD\rightarrow YE'}$ and express this as $\tilde{V}_{AB}\ket{\Psi}=\tilde{V}_{CD}^T\ket{\Psi}$. 
Repeating the same reasoning for several other possible pairings yields
\begin{align}
    \tilde{V}_{AB}\ket{\Psi}=\tilde{V}_{CD}^T\ket{\Psi} = \tilde{V}_{EA}\ket{\Psi} =\tilde{V}_{BC}^T\ket{\Psi} = \tilde{V}_{DE}\ket{\Psi} = \tilde{V}_{AB}^T\ket{\Psi}.
\end{align}
But there is no non-trivial state such that $\tilde{V}_{AB}\ket{\Psi}=\tilde{V}_{AB}^T\ket{\Psi}$ for all $V$, as we show in \cref{lem:transpose}. 
This contradiction shows that the $((2,2,5))$ scheme cannot be realized in the unknown partner setting. 
In the main text (\cref{sec:neces-ESS-unknown}) we provide further necessary conditions on realizable pair access structures in this case.

\subsection{Related work}

As mentioned above, our work is inspired by the definition of quantum secret sharing (QSS).
In fact, ESS contains QSS as a special case, in a sense we can make precise. 
Consider an ESS with known partners.
Suppose the QSS has authorized sets $\mathcal{A}_{QSS}=\{A_1,...,A_k\}$ and unauthorized sets $\mathcal{U}_{QSS}=\{U_1,...,U_\ell\}$. 
Then we take the ESS to have authorized pairs $\mathcal{A}_{ESS}=\{\{S_0,A_1\}, ...,\{S_0,A_k\}\}$ and unauthorized pairs $\mathcal{U}_{ESS}=\{\{S_0,U_1\}, ...,\{S_0,U_\ell\}\}$. 
Since $S_0$ is maximally entangled with $S_1...S_n$, we can view the ESS state as prepared by a channel $\mathcal{N}_{S_{0}'\rightarrow S_1...S_n}$, according to $\Psi_{S_0S_1...S_n}=\mathcal{N}_{S_0'\rightarrow S_1...S_n}({\Psi^+}_{S_0S_0'})$.
It then follows that $\mathcal{N}$ defines an encoding map with authorized sets $\mathcal{A}_{QSS}=\{A_1,...,A_k\}$ as needed.
Additionally, the sets $\mathcal{U}=\{U_1,...,U_\ell\}$ will not be able to recover the secret, though they may obtain partial information.
If we require in our definition of ESS that unauthorized pairs be product, then we recover exactly the security requirement of QSS. 
Note that normally we take a more relaxed requirement, wherein unauthorized pairs cannot recover a state within a fixed distance of the maximally entangled state. 

In \cite{choi2013entanglement}, the authors define what they call entanglement sharing schemes, which we claimed above we generalize.
More precisely, taking our requirement on unauthorized pairs to be that they hold a separable state, we recover the earlier notion of ESS as a special case. 
In their setting, they consider a dealer who retains one end of a maximally entangled state, call it $S_0$. 
The other end of the maximally entangled state is encoded into a set of $n$ shares. 
A subset in a list of authorized sets $\{A_1,...,A_k\}$ should then be able to recover entanglement with $S_0$, while a subset in a second list of unauthorized sets $\{U_1,...,U_\ell\}$ should be in a separable state with $S_0$. 
This is immediately constructed by our notion of ESS if we take authorized pairs $\mathcal{A}_{ESS}=\{\{S_0,A_1\}, ...,\{S_0,A_k\}\}$ and unauthorized pairs $\mathcal{U}_{ESS}=\{\{S_0,U_1\}, ...,\{S_0,U_\ell\}\}$, and require unauthorized pairs be separable. 
Our definition generalizes this setting by not having a privileged share $S_0$. 

Another setting with some similarities to ESS is the construction of \emph{absolutely maximally entangled states} \cite{helwig2012absolute}. 
In that setting, we consider $n$ players each holding a qudit of dimension $d$. 
We then try to construct states on the $n$-partite Hilbert space shared by the players with the property that across any bipartition into sets of size $(m,n-m)$ with $m\leq n/2$, the $m$-qudit system is maximally entangled with the larger $n-m$ qudit system. 
This is known as an AME$(n,d)$ state.
Some of our constructions of ESS are also AME states. 
For instance, consider the state in \cref{eq:225}. 
A brief calculation shows that any two subsystems are maximally entangled with the remaining three, so this is an AME$(5,5)$ state.
In general, one can view an AME state as an ESS with very special authorized pairs and an empty set of unauthorized pairs; meanwhile most ESS cannot be constructed from AME states.   

Outside of specific constructions, our general definition of ESS does not encapsulate the notion of AME states. 
For instance, in AME states some systems are required to be maximally entangled with others.
We never place such a requirement, instead asking that a fixed size EPR pair can be recovered from certain pairs of subsystems, and then we allow the local subsystem dimension to be chosen as part of the construction.
Thus our local subsystems are not required to be maximally entangled. 
Interestingly however, for at least one of the applications of AME states proposed in \cite{helwig2012absolute} ESS seems to be a more powerful tool, which allows for novel functionality.
In the AME context \cite{helwig2012absolute} proposes a many party teleportation protocol wherein the state is distributed, and then later a bipartition of the parties is chosen. 
The parties recover entanglement from the AME state and use this to perform teleportation. 
Using ESS, we can relax the requirement on the sets of parties performing the teleportation from a bipartition, to allowing some shares of the state to be lost. 
For instance using a $((2,2,5))$ scheme, any one share can be lost, and sets of 2 parties each can use the remaining shares to teleport to one another. 

One key distinction between our work and earlier constructions of AME states and related ideas \cite{rico2026threshold,goyeneche2015absolutely,raissi2020constructions} is that we do not require our subsystems to play a symmetric role in the overall entanglement of the state.
Instead, our framework allows us to construct more flexible entanglement structures in our states.
In future work it may be interesting to combine this flexible entanglement aspect of ESS with the maximal entanglement requirement in AME states, by considering the construction of quantum states where only certain subsystems must be maximally entangled.

Another similar line of work is initiated in \cite{rico2026threshold}, where the authors study what they call \emph{threshold entanglement sharing}.
These are pure quantum states such that pairs of sufficiently small subsystems are absolutely separable.\footnote{A density matrix $\rho_{S_1...S_k}$ is said to be absolutely separable if $U_{S_1...S_k}\rho_{S_1...S_k}U^\dagger_{S_1...S_k}$ is separable across every bipartition, for every unitary $U$.}
This is similar to our condition requiring some pairs of subsets be unauthorized, and in particular (for one notion of security we consider) separable. 
Again however, we consider a more asymmetric setting, where a specific list of pairs of subsets should be separable, rather than all pairs of certain sizes.
As well, we only require these pairs be separable across the associated bipartition, rather than that their density matrix be absolutely separable. 
Further, we combine the separability condition with a simultaneous entanglement condition via our authorized pairs. 

\subsection{Our results}

We study ESS in the known and unknown partner settings; we also study in detail certain special cases of both settings. 
We briefly summarize our results here.

Beginning with the known partner case (\cref{sec:ESSknown}), we prove the following condition is necessary.
\begin{restatable}{lemma}{knownpartnermonotonicity}\label{lemma:ESSmonotonicity} \textbf{(monotonicity)}
The access structure for an entanglement sharing scheme with a known partner must satisfy monotonicity. 
\end{restatable}
Here monotonicity means that if $A=\{T_1,T_2\}$ is authorized and $T_1\subseteq T_1', T_2\subseteq T_2'$, then $A'=\{T_1',T_2'\}$ is also authorized. 
In fact, we conjecture that this condition is also sufficient. 
Towards this, we give a construction based on Haar random unitary encodings that we argue achieves a weakened notion of security for ESS schemes.\footnote{In particular, our candidate construction may allow recovery of a state arbitrarily close to (but not exactly) a maximally entangled state from unauthorized pairs. }
See \cref{sec:knownpartnersketch}. 

Continuing from the example given in \cref{sec:firstexample}, we construct $((a,b,a+2b-1))$ threshold entanglement sharing schemes, or, by throwing away some shares $((a,b,s))$ schemes with $a+b\leq s\leq a+2b-1$. 
Our most general construction is based on Reed–Solomon codes, which generalizes the classical secret sharing based construction of the $((2,2,5))$ scheme.

We also give a complete characterization of known partner ESS when restricting to stabilizer encodings. 
Theoretically, ESS and QSS are combinatorial objects closely related to quantum error correcting codes (QECC), and stabilizer codes are the most important and well-studied QECCs, thus stabilizer encodings are a natural focus for this work, giving us a useful toolkit and expanding potential applications.  
Practically, stabilizer states and operations are easier to prepare and to perform, with many features helpful for fighting noise and achieving fault-tolerance.
So, it is of great practical interest to understand how much can be achieved within this setting. 
As well, we find it intriguing that stabilizer encodings can realize only a subset of all possible access structures; this is in contrast to the case of quantum secret sharing, where stabilizer codes suffice to realize every allowed access structure. 
In the stabilizer case we find that there are two conditions which are necessary and sufficient:
\begin{itemize}
    \item Monotonicity, as in \cref{lemma:ESSmonotonicity}. 
    \item A second condition we call the rank condition, stated as \cref{lemma:rank}, which expresses that authorized pairs in stabilizer encodings always come with additional authorized pairs formed by repartitioning the shares involved.
\end{itemize} 
To understand the rank condition we consider an example with three shares, $\{S_1, S_2, S_3\}$. 
Consider the access structure with only one authorized pair $\{\{S_1\},\{S_2,S_3\}\}$ (all other pairings are unauthorized). 
Then, the rank condition says that this access structure cannot be realized with a stabilizer encoding. 
Instead, if $\{\{S_1\},\{S_2,S_3\}\}$ is authorized, then, at least one of $\{\{S_1,S_2\},\{S_3\}\}$ or $\{\{S_1,S_3\},\{S_2\}\}$ must also be authorized.
In contrast, non-stabilizer encodings lack this repartitioning property; indeed we construct an explicit state realizing only the authorized pair $\{\{S_1\},\{S_2,S_3\}\}$ by using a non-stabilizer encoding (see \cref{sec:knownpartnersketch}).

Moving on to the unknown partner case, we find a much richer set of necessary conditions. 
These conditions are most straightforward to state in terms of the authorized pair graph, labelled $G_{\mathcal{A}}$, which represents each subset $T_i$ as a vertex, and each authorized pair $\{T_i,T_j\}$ as an edge between the vertices $i$ and $j$. 
In this language we can state the following conditions, which we prove are necessary:
\begin{itemize}
    \item \textbf{No odd cycles:} The authorized pair graph must not contain any odd cycles. 
    \item \textbf{Monogamy:} If there is an even length path $\{T_1,T_2\}$,...,$\{T_{k-1},T_{k}\}\in G_{\mathcal{A}}$ ($k$ odd), then $T_1\cap T_{k}\neq\emptyset$.
    \item \textbf{Transitivity:} If there is a path $\{T_1,T_2\}$,...,$\{T_{k-1},T_{k}\}\in G_{\mathcal{A}}$ with odd length and $T_1\cap T_{k}=\emptyset$, then $\{T_1,T_{k}\}$ is authorized.
    \item \textbf{Weak monotonicity:} If $\mathcal{A}$ is maximal,\footnote{A maximal access structure is one that cannot be extended to a larger one. See \cref{def:maximality}.} $\{T_1, T_2\}\in \mathcal{A}$, $T_1\subseteq T_3$ and $T_3\cap T_2=\emptyset$, then 
     $T_3$ occurs in at least one authorized pairing of $\mathcal{A}$.
\end{itemize}
We also argue that these conditions suffice. 
Similarly to the known partner case, we give a candidate construction using Haar random encodings, which we argue realizes any unknown partner access structure satisfying the above conditions. 
However, our candidate construction is only argued to satisfy the same weakened notion of security as in the known partner case. 

We are able to more rigorously understand sufficiency in some restricted settings. 
In \cref{sec-ESSunknownsimple} we study the case with only authorized pairs, and no unauthorized pairs, where we find a complete characterization: we prove that the no odd cycles and monogamy conditions are necessary and sufficient in that case. 
Further, in \cref{sec-stabiffunknown} we give a complete characterization of the stabilizer case.
Here, the no odd cycles condition and transitivity are maintained, while the remaining conditions are replaced by a set of linear constraints on a matrix $u$, which has $0,1$ entries recording the commutation relations of the stabilizers. 

In \cref{sec:lowerbounds}, we give some simple lower bounds on the size of shares $S_i$ needed to construct a given access structure. 
In \cref{sec:summoning}, we give an application of ESS to entanglement summoning. 

\section{Background and preliminaries} \label{sec:background}

\subsection{Quantum information tools}\label{sec:QIbasics}

We define the fidelity of two density matrices $\rho,\sigma$ as
\begin{align}
    F(\rho,\sigma) = \left(\tr \sqrt{\sqrt{\rho}\sigma\sqrt{\rho}}\right)^2.
\end{align}
We define the von Neumann entropy of a density matrix $\rho_A$ as 
\begin{align}
    S(A)_\rho = -\tr (\rho_A \log \rho_A).
\end{align}
The conditional quantum mutual information is 
\begin{align}
    I(A:C|B)_\rho = S(AB)_\rho+S(BC)_\rho-S(B)_\rho-S(ABC)_\rho.
\end{align}
The von Neumann entropy satisfies strong subadditivity, which means
\begin{align}
    I(A:C|B)_\rho\geq 0.
\end{align}

The squashed entanglement \cite{christandl2004squashed,brandao2011faithful} is defined as 
\begin{equation}
    E_{\text{sq}}(A:B)_\rho = \inf_{\sigma_{ABC}: \tr_C(\sigma)=\rho} \frac{1}{2}I(A:B|C).
\end{equation}
It satisfies:
\begin{equation}
    E_D(A:B)_\rho \leq E_{\text{sq}}(A:B)_\rho \leq \frac{1}{2}I(A:B)_\rho\leq S(A)_\rho,
\end{equation}
where $E_D$ is the distillable entanglement.
Moreover, the squashed entanglement satisfies the monogamy inequality \cite{koashi2004monogamy}: for any $\rho_{ABC}$,
\begin{equation}
    E_{\text{sq}}(A:B)_\rho+ E_{\text{sq}}(A:C)_\rho \leq E_{\text{sq}}(A:BC)_\rho.
\end{equation}
The relative entropy of entanglement is defined as
\begin{align}
    E_R(A:B)_\rho = \min_{\sigma_{AB}\in \text{SEP}_{AB}} D(\rho_{AB}||\sigma_{AB}),
\end{align}
where $D(\rho||\sigma)=\tr(\rho\log\rho-\rho\log\sigma)$ is the quantum relative entropy and $\text{SEP}_{AB}$ is the set of separable states on $AB$.
The relative entropy of entanglement satisfies the following continuity property \cite{horodecki2005locking}, 
\begin{align}\label{eq:SRcontinuity}
    E_R(AA':B)_\rho\leq E_R(A':B)_\rho + 2\log d_A.
\end{align}

\subsection{Quantum secret sharing}\label{sec:QSS}

The setting we study can be viewed as a bipartite analogue of the traditional framework of quantum secret sharing, which also proves to be a useful tool in the bipartite case. 
For clarity, we briefly recall the definition of quantum secret sharing and some basic facts about access structures.

\begin{definition}
        Consider a set $\Gamma = \{S_1, \cdots, S_n\}$ of $n$ parties. 
        An \textbf{access structure} $\mathcal{S}=(\mathcal{A}, \mathcal{U})$ on $\Gamma$ consists of an authorized set $\mathcal{A}=\{A_i: A_i \subset \Gamma\}_i$ and an unauthorized set $\mathcal{U}=\{U_{\ell}:U_{\ell} \subset \Gamma\}_{\ell}$. Each party $S_j$ holds a system which we denote also by $S_j$ and is called a \textbf{share}. 
\end{definition}

A \textbf{threshold access structure} is a special case of practical interest.
Given an integer $1 \leq k \leq n$, all subsets of $k$ or more parties are required to be authorized, while all subsets of size less than $k$ are required to be unauthorized.
This is denoted as a $((k,n))$ threshold scheme.

Next, we define a quantum secret sharing scheme \cite{cleve1999share}. 
Here and throughout the paper, for a quantum channel $\mathcal{N}_{A\rightarrow BC}$, we denote $\mathcal{N}_{A\rightarrow B} := \tr_C \circ \mathcal{N}_{A\rightarrow BC}$. 
\begin{definition}
    A \textbf{quantum secret sharing scheme} realizing access structure $\mathcal{S}=(\mathcal{A}, \mathcal{U})$ on $n$ parties is an encoding map $\mathcal{E}_{Q\rightarrow S_1...S_n}$ such that
    \begin{itemize}
        \item For all $A_i\in \mathcal{A}$, there exists a decoding map $\mathcal{D}_{A_i\rightarrow Q}$ such that 
        \begin{align}\label{eq:qss1}
            \mathcal{D}_{A_i\rightarrow Q}\circ \mathcal{E}_{Q\rightarrow A_i}=\mathcal{I}_{Q\rightarrow Q},
        \end{align} 
        where $\mathcal{I}$ denotes the identity channel.
        \item For all $U_{\ell}\in \mathcal{U}$, there exists a state preparation channel $P_{\emptyset \rightarrow U_{\ell}}$ such that 
        \begin{align}\label{eq:qss2}
        \mathcal{E}_{Q\rightarrow U_{\ell}}= P_{\emptyset \rightarrow U_{\ell}} \circ {\tr}_{Q}.
        \end{align}
    \end{itemize}
\end{definition}

There are efficient constructions of threshold quantum secret sharing schemes \cite{cleve1999share}, in the sense that the constructions use subsystems with small dimensions.
In fact these constructions use shares whose dimension matches that of the secret, which is known to be optimal.
To see why, we review the following notion of a significant share.
\begin{definition}
A share $S_j$ in an access structure $\mathcal{S}$ is called a \textbf{significant share} if there exists an authorized set $A_i$ and an unauthorized set $U_{\ell}$ such that $U_{\ell} \cup \{S_j\} = A_i$.
\end{definition}
We now state a lemma that constrains the size of significant shares. For completeness, we reproduce the proof here (see also \cite{gottesman2000theory,imai2003quantum}).
\begin{lemma} Let $S_j$ be a significant share in a secret sharing scheme, and its dimension be $d_{S_j}$. Further let the secret be of dimension $d_Q$. Then $d_{S_j}\geq d_{Q}$. In other words, any significant share must have dimension at least that of the secret. 
\end{lemma}
\begin{proof}\,
    Consider preparing the system $Q$ in a maximally entangled state with reference $R$. 
    First, notice that for any authorized set, in particular $A_i$, we have
    \begin{align}
        I(R:A_i)= 2n_Q
    \end{align}
    where $n_Q=\log \dim Q$. 
    This follows from \cref{eq:qss1} along with the data processing inequality: there is a decoder acting on $A_i$ which recovers the maximally entangled state with $R$ from $A_i$, so $A_i$ already must be maximally correlated with $R$.
    Then, notice that 
    \begin{align}
        I(R:U_{\ell})=0
    \end{align}
    since \cref{eq:qss2} requires $R$ and $U_{\ell}$ form a product state. 
    Then using the entropy inequality, 
    \begin{align} \label{eq:lastlinesignshare}
        I(R:AB) \leq I(R:A) + 2S(B) \leq I(R:A)+2\log \dim B
    \end{align}
    with $B=S_j$, $A=U_{\ell}$, we are done. 
\end{proof}

We call an access structure \emph{valid} if there exists a secret sharing scheme that realizes it.
Not every access structure is valid. Two necessary conditions are:
\begin{itemize}
\item \textbf{Monotonicity:} If $A \subseteq X$ and $A$ is authorized, then $X$ must also be authorized.
\item \textbf{No-cloning:} If $A$ and $A'$ are both authorized, then $A \cap A' \neq \emptyset$.
\end{itemize}

The first condition is immediate: if recovery is possible from a smaller set, then it is also possible from any larger set by, for instance, discarding the additional systems and applying the recovery channel for the smaller set.
The second condition follows directly from the no-cloning theorem: if two disjoint authorized sets existed, each could independently reconstruct the secret, thereby producing two copies.

Remarkably, these two conditions are not only necessary but also sufficient for an access structure to be valid.
Explicit constructions of secret sharing schemes for arbitrary access structures can be found in \cite{gottesman2000theory, smith2000quantum}.
We also provide in \cref{app-QSS} an alternative construction that establishes sufficiency.

In a \emph{classical secret sharing scheme} \cite{shamir1979share}, a secret $s$ (consisting of a classical bit string) is encoded into strings $m_i\in S_i$, with $S_i$ random variables. 
Each of $n$ players receives one of the random variables $S_i$. 
The encoding involves a map that also takes in classical randomness $r$, 
\begin{align}
    E(s,r)=(m_1,...,m_n).
\end{align}
The players are not given access to the randomness $r$. 
Similar to the quantum case, we define authorized and unauthorized subsets of the shares, which together define the access structure for the scheme. 
Classically, the monotonicity condition is necessary and sufficient for a scheme to be realizable \cite{ito1989secret}. 

A classical $(k,n)$ threshold secret sharing scheme involves $n$ shares such that any $k$ shares can recover the secret, but all subsets of size $k-1$ cannot. 

\subsection{Pair access structures}\label{sec:pairaccess}

A (quantum) secret sharing scheme is characterized by a collection of authorized sets and a collection of unauthorized sets. 
In the bipartite case, we need a somewhat more complicated object to characterize an access structure. 
\begin{definition}
    Consider a set $\Gamma = \{S_1, \cdots, S_n\}$ of $n$ parties. A \textbf{pair access structure} $\mathcal{S}=(\mathcal{A}, \mathcal{U})$ on $n$ parties consists of a set of authorized pairs $\mathcal{A}=\{A_k\}$, and a set of unauthorized pairs $\mathcal{U}=\{U_\ell\}$.
    Each $A_k$ or $U_\ell$ is an unordered pair of two subsets, $\{T_i, T_j\}$, with $T_i$ and $T_j$ subsets of $\Gamma$. We require that any pairing $\{T_i,T_j\}$ occurring as an authorized or unauthorized set be disjoint: $T_i\cap T_j =\emptyset$.
    We also require that $\mathcal{A}\cap \mathcal{U}=\emptyset$, i.e. no set is deemed both authorized and unauthorized. 
\end{definition}
Further, we introduce a partial ordering on pairs as follows. 
\begin{definition}
    Given two pairs $X=\{T_1,T_2\}$, $Y=\{T_3,T_4\}$, we say $X\preceq Y$ if at least one of the following is true:
    \begin{itemize}
        \item $T_1\subseteq T_3$ and $T_2\subseteq T_4$
        \item $T_1\subseteq T_4$ and $T_2\subseteq T_3$
    \end{itemize}
    We say $X\prec Y$ if $X\preceq Y$ and $X\neq Y$. 
\end{definition}
With this partial ordering in hand, we can give a definition of monotonicity of a pair access structure. 
\begin{definition}\label{def:monotonicity}
    We say that an access structure $\mathcal{A}$ satisfies \textbf{monotonicity} if for all pairs $A$ and $A'$, whenever $A\in \mathcal{A}$ and $A\preceq A'$ we also have $A'\in \mathcal{A}$. 
\end{definition}

A special case of a pair access structure is a threshold structure, which has the property that all pairs which are large enough are authorized. 
To define this precisely, we introduce the following notion. 
\begin{definition}
    A pairing $\{T_i,T_j\}$ is said to have \textbf{type} $\{a,b\}$ if $\{|T_i|,|T_j|\}=\{a,b\}$. 
\end{definition}
Now we can define a threshold pair access structure. 
\begin{definition}
    A pair access structure $(\mathcal{A}, \mathcal{U})$ is said to be a type-$((a,b,n))$ \textbf{threshold structure} if it is a pair access structure on $n$ parties such that:
    \begin{itemize}
        \item All pairs of type $\{a,b\}$ are authorized.
        \item All pairs $J$ such that there exists a pairing $K$ of type $\{a,b\}$ with $K\preceq J$ are authorized. 
        \item All pairs $J$ such that there exists a $K$ of type $\{a,b\}$ with $J\prec K$ are unauthorized. 
    \end{itemize} 
\end{definition}
As an example, a $((2,2,5))$ pair access structure has all pairings of type $\{2,2\}$ and $\{2,3\}$ authorized, and all pairings of type $\{1,1\}$, $\{1,2\}$ unauthorized.
Pairings of type $\{1,3\}$ and $\{1,4\}$ are incomparable to any of type $\{2,2\}$, so the structure does not place a requirement on them (they may be authorized or unauthorized). 

We also use a few more definitions related to pair access structures. 
\begin{definition}\label{def:flatten}
    Given a set of authorized pairs $\A$, its \textbf{flattening} $\cup\A$ is defined as:
\begin{equation}
    \cup\A = \{ T\,|\, \exists \,T' \,\text{s.t. } \{T, T'\}\in \A\}.
\end{equation}
\end{definition}
In other words, we collect all subsets (of the party set) $T$ such that $T$ occurs in an authorized pair.

In some settings it will be helpful to view the authorized sets of an access structure as a graph, with the subsets $T_i$ labelling the vertices and edges added for each authorized pairing. 
\begin{definition}\label{def-graph}
    The \textbf{authorized pair graph} $G_\mathcal{A}$ is the undirected graph with vertex set $\cup\A$, and edge set $\A$: 
    it includes a vertex $v_i$ for each subset $T_i$ that occurs in an $A\in \mathcal{A}$, and an edge $\{v_i,v_j\}$ for each authorized pair $\{T_i,T_j\}\in \mathcal{A}$.
\end{definition}
An example is shown as \cref{fig:examplepairgraph}.

\begin{figure}
\centering
\begin{tikzpicture}
  
\draw[fill=black] (0,2) circle (0.05cm);
\node[above] at (0,2) {$T_1$};

\draw[thick] (0,2) -- (-1.25,0);
\draw[thick] (0,2) -- (1.25,0);

\draw[fill=black] (-1.25,0) circle (0.05cm);
\node[below] at (-1.25,0) {$T_2$};

\draw[fill=black] (1.25,0) circle (0.05cm);
\node[below] at (1.25,0) {$T_3$};

\end{tikzpicture}
\caption{A simple example of an authorized pair graph (\cref{def-graph}). The graph indicates that $\{T_1,T_2\}$ and $\{T_1,T_3\}$ are authorized, but $\{T_2,T_3\}$ is not.}
\label{fig:examplepairgraph}
\end{figure}

For later use we record the following simple lemma from graph theory. 

\begin{lemma}\label{lemma:oddcyclesandbipartite}
    A graph has no odd cycles if and only if it is bipartite. 
\end{lemma}

\subsection{Stabilizer codes}\label{sec:stabilizerstuff}

Stabilizer codes \cite{gottesman1996,GF41998} are a quantum generalization of classical linear codes. Most of the quantum error correcting codes known to date are stabilizer codes. 
Furthermore, the stabilizer formalism provides a succinct method to track quantum evolution.
Interested readers can consult introductory treatments such as \cite{NielsenChuang2010,PreskillQEC2026}, or comprehensive treatments in \cite{gottesmanthesis1997,gottesmanbook2026}. 
Here, we review some useful facts about stabilizer codes, as well as the construction of a specific class of stabilizer codes known as Reed–Solomon codes \cite{AB1997,GGB1999l}. 
Then, we derive a useful expression for the density matrix for a stabilizer state as a sum of certain stabilizers.  
At the end of this subsection, we will state a powerful structural theorem for the entanglement in any tripartite stabilizer state \cite{Griffiths-Looi-2011,BFG2006}, and give a method of revealing if an EPR pair can be distilled from a stabilizer state based on the commutation properties of its stabilizers.    

We consider quantum systems consisting of $n$ qudits, each of dimension $p$ which is a prime number.
In this setting, $X$ and $Z$ operators are defined by
\begin{align}
    X\ket{j}=\ket{j+1},~~
    Z\ket{j}=\omega^j\ket{j},
\end{align}
where $\omega=e^{i2\pi/p}$ is a $p$-th root of unity, and these operators satisfy $ZX=\omega XZ$.

The quantum stabilizer codes generalize classical linear codes in the following way. 
A classical linear code is defined by all codewords whose entries satisfy some linear constraints. 
For instance, the $3$-bit binary repetition code has two codewords $000$ and $111$, which can also be specified by the linear constraint of requiring the first two entries sum to an even number, and the last two entries sum to an even number. 
More generally, the linear constraints on strings $x$ can be recorded into a matrix $H$, called the parity check matrix, with the valid code words being those in the kernel of $H$.
Moving to the quantum setting, linear constraints can be specified by restricting to states in the eigenspace of a chosen Hermitian operator.
A quantum stabilizer code is defined as the subspace of the $n$-qudit system that is the simultaneous $+1$ eigenspace of a list of mutually commuting tensor-product Pauli operators.  
More precisely, we begin with a set of mutually commuting Pauli operators that generate (multiplicatively) an Abelian group $S$, called the stabilizer group, such that $e^{i\phi}I \notin S$ for any $e^{i\phi} \neq 1$.
The common $+1$ eigenspace of all stabilizers $g \in S$ is called the stabilizer code associated with $S$.
The corresponding stabilizer state $\rho$ is defined as the maximally mixed state on the code space. 

We will use a special class of stabilizer codes called Calderbank–Shor–Steane (CSS) codes \cite{CalderbankShor1996,Steane1996}.
It is special in that each generator of the stabilizer group can be chosen to be a tensor product of $X$ and $I$, or of $Z$ and $I$. Operationally, the $X$-type generators correct the $Z$ errors, and the $Z$-type generators correct the $X$ errors. 
These two types of error correction are specified by two classical linear codes $C_1$, $C_2$, with parity check matrices $H_1$ and $H_2$ respectively. 
The $Z$-type and $X$-type
stabilizer generators are defined as:
\begin{align}
S^Z_i=Z(H_{1,i})=\bigotimes_{j=1}^n(Z_j)^{H_{1,ij}}, \\
S^X_i=X(H_{2,i})=\bigotimes_{j=1}^n(X_j)^{H_{2,ij}},
\end{align}
where the $H_{a,i}$ in the middle denotes the $i$-th row of $H_a$; $Z_j$ and $X_j$ denote the Pauli $Z$ and $X$ matrices on qudit $j$, respectively.
To ensure that 
the $Z$- and $X$-type stabilizers commute, 
the parity check matrices of the classical codes must satisfy the relation $H_1H_2^T=0$, which is equivalent to $C_2^\perp\subseteq C_1$ where $C_2^\perp$ is the orthogonal complement of $C_2$. 
Conversely, a collection of mutually commuting $X$- and $Z$-type stabilizers gives rise to a CSS code. 
A CSS code is specified by the following symplectic representation via a matrix over $\mathbb{F}_p$:
\begin{align}\label{eq:symplectic}
    \left(\begin{array}{c|c}
0 & H_1 \\
H_2 & 0 
\end{array}\right).
\end{align}

The dimension of the code subspace is given by $|C_1|/|C_2^\perp|$, with a basis given by the following coset states:
\begin{align}\label{eq:CSSstate}
    \ket{x+C_2^\perp} = \frac{1}{\sqrt{|C_2^\perp|}}\sum_{y\in C_2^\perp} \ket{x+y},
\end{align}
where $x\in C_1$.
To see that they are indeed invariant under $S^Z_i$ and $S^X_i$ defined above, 
note that on the RHS, each summand is a codeword of $C_1$, hence invariant under $S_i^Z$.
Meanwhile, $S_i^X$ shifts each $\ket{x}$ to $\ket{x+y_0}$ where $y_0$ is a fixed element in $C_2^\perp$, hence the summation is invariant under $S_i^X$.

We now describe two types of logical operators for a CSS code: $Z$-type and $X$-type. 
For each classical codeword $\bar{z}\in C_2 \setminus C_1^\perp$ and $\bar{x}\in C_1 \setminus C_2^\perp$, define:
\begin{align}
    L_Z &= Z(\bar{z}) = \bigotimes_{j=1}^n (Z_j)^{\bar{z}_j}, \\
    L_X &= X(\bar{x}) = \bigotimes_{j=1}^n (X_j)^{\bar{x}_j}.
\end{align}
If we apply $L_Z$ to the states defined in \cref{eq:CSSstate}, because $\bar{z}\in C_2$, $L_Z$ is unaffected by the $y$ term in the summand, but since $\bar{z} \notin C_1^\perp$, $L_Z$ will incur a non-trivial phase for some $x \in C_1$. Thus $L_Z$ acts like a logical $Z$ on states defined in \cref{eq:CSSstate}.
If we apply $L_X$ instead, since $\bar{x}\in C_1 \setminus C_2^\perp$, each summand becomes $\ket{x+y+\bar{x}}$ 
with $x+y+\bar{x} \in C_1$ but since $\bar{x}\notin C_2^\perp$, $x+C_2^\perp \neq \bar{x}+x+C_2^\perp$, hence, $L_X$ acts like a logical $X$ operator. 

We provide an example to illustrate the above ideas. 
Consider the first example in \cref{sec:firstexample}, the state shown in \cref{eq:225}. 
The stabilizer group is generated by
\begin{align}
    XXXXX,\, IXX^2X^3X^4,\,I Z^{\dagger} ZZ Z^{\dagger},\,Z^{\dagger} I Z^{\dagger} ZZ,\,Z Z^{\dagger} I Z^{\dagger} Z. 
\end{align}
The $XXXXX$ operator can be seen to stabilize the state in \cref{sec:firstexample} by re-indexing the sum over $k$; the $IXX^2X^3X^4$ operator can be seen to stabilize the same state by re-indexing the sum over $s$.
Note that $Z^{\dagger} Z Z Z^{\dagger} I$ and $Z Z Z^{\dagger} I Z^{\dagger}$ are also stabilizers but are not independent of the five already listed. 
Note also that the generators indeed commute with one another. 
We have 
\[H_1 = \left[ \begin{array}{ccccc} 
0 &4 &1 &1 &4 \\
4 &0 &4 &1 &1 \\
1 &4 &0 &4 &1 \end{array} \right], ~~~~
H_2 = \left[ \begin{array}{ccccc} 
1 &1 &1 &1 &1 \\
0 &1 &2 &3 & 4\end{array} \right]
\]
and $H_1 H_2^T = 0$. Since there are 5 stabilizers and $5$ qudits of dimension $5$, the code space is 1-dimensional and there are no nontrivial logical operators. 

In this paper we frequently use a special class of CSS codes called the quantum Reed–Solomon codes \cite{AB1997,GGB1999l}. Here we record only the definitions and properties that will be needed; a more detailed account, including proofs, is provided in \cref{sec:RScodes}.

For $p\geq n$, an $[[n,k]]_p$ quantum Reed–Solomon code is specified by $n$ distinct elements $x_0, \cdots, x_{n-1}\in \mathbb{F}_p$ and a parameter $r$ such that $1\leq r\leq n-k$.
Define a family of polynomials $f_{c,s}(x)$, where $c\in \mathbb{F}_p^r$ and $s\in \mathbb{F}_p^k$, as follows:
\begin{equation}\label{eq:RSpoly}
    f_{c,s}(x)=c_0+c_1x+\cdots+c_{r-1}x^{r-1}+s_0x^r+\cdots+s_{k-1}x^{r+k-1}.
\end{equation}
The code states are then defined as:
\begin{equation}\label{eq:RSwavefunction}
    \ket{\bar{s}}=\frac{1}{p^{r/2}}\sum_{c\in \mathbb{F}_p^r}\ket{f_{c,s}(x_0)}\otimes \ket{f_{c,s}(x_1)}\cdots\otimes\ket{f_{c,s}(x_{n-1})}.
\end{equation}
For our purposes, we need the following result on the distance\footnote{The distance of a code is the minimal number of qudits that an operator needs to act on to give a (non-trivial) logical operator.} of this code.
\begin{lemma}\label{lemma:RScode}
For the $[[n,k]]_p$ quantum Reed–Solomon code defined above, 
any set of $r+1$ qudits supports a nontrivial $Z$-type logical operator;  no set of $r$ qudits supports a $Z$-type logical operator or stabilizer operator.
Similarly, any set of $n-(r+k)+1$ qudits supports a nontrivial $X$-type logical operator; no set of $n-(r+k)$ qudits supports an $X$-type logical operator or stabilizer operator.
\end{lemma}
\begin{proof}\,
    See \cref{lem:RSZ} and \cref{lemma:RSLX} in \cref{sec:RScodes}.
\end{proof}

In particular, the code has the following code distance:
\begin{equation}
    d=\min\{d_X,d_Z\},~~d_Z=r+1,~~d_X=n-r-k+1.
\end{equation}

A (pure) stabilizer state is a stabilizer code with a one dimensional code space.  Equivalently, for the ambient space $(\mathbb{C}^p)^{\otimes n}$, the stabilizer has $n$ independent generators $G_1,G_2,\cdots,G_n$.  The density matrix for the state can be expressed in terms of the generators: 
{\small \begin{align} \label{eq:purestabilizerstate1} 
\left( \frac{I+G_1+\cdots+G_1^{p-1}}{p} \right) 
\left( \frac{I+G_2+\cdots+G_2^{p-1}}{p} \right) \cdots
\left( \frac{I+G_n+\cdots+G_n^{p-1}}{p} \right) \,.
\end{align}
} $\!\!$To see this, note that the $i$-th factor is the projector onto the $+1$ eigenspace of $G_i$, and the product of these commuting projectors projects onto the simultaneous $+1$ eigenspace.  
The above operator has trace $1$.  Finally, the $G_i$'s are independent, so the operator has rank $1$, and indeed (\ref{eq:purestabilizerstate1}) gives the density matrix of the stabilizer state.   
Now if we multiply out the terms in each factor, we have $p^n$ terms, each being a unique product of the $G_i's$.  So, those $p^n$ terms are precisely the elements of the stabilizer group generated by the $G_i's$.  This gives a second expression for the density matrix $\rho$ of a pure stabilizer state whose stabilizer group is $S$: 
\begin{align} \label{eq:purestabilizerstate2} 
\rho = \frac{1}{p^n} \sum_{g \in S} g.
\end{align}

We can derive expressions analogous to (\ref{eq:purestabilizerstate1}) and (\ref{eq:purestabilizerstate2}) in two additional settings when the state is mixed.  
In the first setting, we have only $l < n$ generators for the stabilizer, and the mixed state is taken to be 
{\small \begin{align} \label{eq:mixedstabilizerstate1} 
\!\!\!\! {1 \over p^{n-l}} \! \left( \frac{I+G_1+\cdots+G_1^{p-1}}{p} \right) \!\!
\left( \frac{I+G_2+\cdots+G_2^{p-1}}{p} \right) \! \cdots \!
\left( \frac{I+G_l+\cdots+G_l^{p-1}}{p} \right) .
\end{align}
} $\!\!$This is the usual setting for stabilizer codes, and the above operator is the maximally mixed state on the code space (and proportional to the projector onto the code space). 
The expression (\ref{eq:purestabilizerstate2}) still holds in this case but the group $S$ has fewer elements.
In the second setting, some of the $n$ qudits are traced out. Let $A$ be the set of the remaining qudits, and $B$ be the set of qudits being traced out. 
The reduced state on $A$ can be readily obtained from (\ref{eq:purestabilizerstate2}). 
All non-trivial Pauli operators are traceless, so, as we trace out $B$, the only terms remaining are those with identity on $B$, which are also the ones with support contained in $A$.
Thus, 
\begin{align} \label{eq:mixedstabilizerstate2} 
\rho = \frac{1}{p^{|A|}} \sum_{g \in S, \supp(g)\subseteq A} g \,.
\end{align}

We now state a powerful structural theorem for the entanglement in any tripartite stabilizer state \cite{Griffiths-Looi-2011,BFG2006}.

\begin{remark}\label{remark:tripartitestabilizer} For any prime dimension $p \geq 2$, for any stabilizer state $\ket{\psi} \in (\mathbb{C}^p)^{\otimes n}$, for any tripartition $(T_1, T_2, R)$ of the $n$ qudits, and up to local Clifford unitaries on the three parts, $\ket{\psi}$ is a tensor product of single-qudit states, maximally entangled EPR pairs, and tripartite GHZ states. 
Note that if three parties share only a GHZ state, any two of them share a separable state.
Note that the local unitaries relating $\ket{\psi}$ to the single-qudit, EPR, and GHZ states can depend on the tripartition $(T_1,T_2,R)$.
\end{remark}

Finally, we give a characterization of how to identify when a stabilizer state contains a maximally entangled state across two subsystems. 

\begin{lemma}\label{lemma:stabdistillcondition}
    Suppose that a stabilizer state $\psi_{T_1T_2R}$ has stabilizers $A=A_{T_1}\otimes A_{T_2}\otimes I_R$ and $B=B_{T_1}\otimes B_{T_2}\otimes I_R$, where, for $\ell\neq 0$,
    \begin{align}
        B_{T_1}A_{T_1}=\omega^\ell A_{T_1}B_{T_1}, \quad B_{T_2}A_{T_2}=\omega^{-\ell} A_{T_2}B_{T_2}.
    \end{align}
    Then, it is possible to recover a maximally entangled state by local operations on $T_1$ and local operations on $T_2$.
    Conversely, if there are no stabilizers of the form $A_{T_1}\otimes A_{T_2}$ and $B_{T_1}\otimes B_{T_2}$ satisfying the above commutation relations, then it is not possible to recover an EPR pair. 
\end{lemma} 
\begin{proof}\,
In the stabilizer formalism, if we transform the state 
$\psi$ to $U{\psi}U^\dagger$, each stabilizer is transformed from $S$ to $USU^\dagger$. 
A Clifford unitary can be specified up to an overall phase by how it conjugates the generators of the Pauli group. 
Conversely, transformations of the generators that preserve the commutation relation of the generators are realizable as conjugation by a Clifford unitary \cite{gottesmanbook2026}. 
By assumption, $A_{T_1}$ and $B_{T_1}$ have the same commutation relation as $X^\ell$ and $Z$ so $U_{T_1}$ exists such that 
\begin{align}
    U_{T_1} A_{T_1} U_{T_1}^\dagger &= X^\ell \otimes I \otimes \cdots \otimes I, \nonumber \\
    U_{T_1} B_{T_1} U_{T_1}^\dagger &= Z \otimes I \otimes \cdots \otimes I.
\end{align}
Similarly, $A_{T_2}$ and $B_{T_2}$ have the same commutation relations as $X^\ell$ and $Z^{-1}$, so there exists $U_{T_2}$ such that  
\begin{align}
    U_{T_2} A_{T_2} U_{T_2}^\dagger &= X^\ell \otimes I \otimes \cdots \otimes I, \nonumber \\
    U_{T_2} B_{T_2} U_{T_2}^\dagger &= Z^{-1} \otimes I \otimes \cdots \otimes I.
\end{align}  
Now, the local unitary $U = U_{T_1} \otimes U_{T_2} \otimes I_R$ conjugates the stabilizers $A_{T_1}\otimes A_{T_2}\otimes I_R$ and $B_{T_1} \otimes B_{T_2}\otimes I_R$ into new stabilizers 
\begin{align}
    &(X^\ell I\cdots I)_{T_1} \otimes (X^\ell I\cdots I)_{T_2} \otimes I_R, \nonumber \\
    &(Z I\cdots I)_{T_1} \otimes (Z^{-1}I\cdots I)_{T_2} \otimes I_R.
\end{align}
Next notice we can replace these two stabilizer generators with 
\begin{align}
    A'=&(X I\cdots I)_{T_1} \otimes (X I\cdots I)_{T_2} \otimes I_R, \nonumber \\
    B'=&(Z I\cdots I)_{T_1} \otimes (Z^{-1}I\cdots I)_{T_2} \otimes I_R,
\end{align}
since we can form $X$ from powers of $X^\ell$, and vice versa. 

Next notice that if another stabilizer generator $G$ has $X^\alpha$ in the first register of $T_1$, it must also have an $X^\alpha$ on the first register of $T_2$ since otherwise it will not commute with $(ZI\cdots I)_{T_1} \otimes (Z^{-1}I\cdots I)_{T_2} \otimes I_R$. 
In the stabilizer formalism, replacing a generator $G$ with $G'G$ (for $G'$ another generator) does not change the stabilizer group. 
So, multiplying the stabilizer generator $G$ by $A^{-\alpha}=(X^{-\alpha} I\cdots I)_{T_1} \otimes (X^{-\alpha} I\cdots I)_{T_2} \otimes I_R$ (which is a stabilizer) removes the $X$'s on the first registers of $T_1$ and $T_2$. 
Similarly, we can remove $Z$'s on the first registers of $T_1$ and $T_2$ for all generators except $A'$ and $B'$.
Repeating this argument for all other generators, all other stabilizers except $A'$ and $B'$ will be the identity on the first qudit registers of $T_1$ and $T_2$. 
Thus the state on first qudits of $T_1$ and $T_2$ is the state stabilized by $X\otimes X$ and $Z\otimes Z^{-1}$, which is exactly the maximally entangled state.

To see the converse claim, suppose that it is possible to recover an EPR pair by local operations. 
Then by \cref{remark:tripartitestabilizer}, there is a way to distill this EPR pair using only local Clifford operations $U_{T_1}\otimes U_{T_2}$. 
But then 
\begin{align}
    A&=(U^\dagger_{T_1}\otimes U^\dagger_{T_2})\left[(XI...I)_{T_1}\otimes (XI...I)_{T_2}\right](U_{T_1}\otimes U_{T_2}), \nonumber \\
    B&=(U^\dagger_{T_1}\otimes U^\dagger_{T_2})\left[(ZI...I)_{T_1}\otimes (Z^\dagger I...I)_{T_2}\right](U_{T_1}\otimes U_{T_2}),
\end{align}
will be stabilizers of $\psi$ with the needed commutation relations and tensor product structure. 
\end{proof}

\section{Entanglement sharing schemes with a known partner}\label{sec:ESSknown}

We first consider the setting with known partners.
Namely, given a pairing $\{T_i,T_j\}$, the party holding $T_i$ knows they are trying to prepare a state shared with $T_j$, and vice versa. 
We focus on the case where the to-be-recovered state is a maximally entangled state, the EPR state, and we use $\Psi^+$ to label its density matrix. 

\subsection{Definition and first example}

We define our setting more precisely below. 

\begin{definition}
    We say a state $\Psi_{S_1...S_n}$ is a $\delta$-secure \textbf{entanglement sharing scheme (ESS) with a known partner} and with pair access structure $\mathcal{S}=(\mathcal{A}, \mathcal{U})$ if:
    \begin{enumerate}
        \item For each $A=\{T_i,T_j\}\in \mathcal{A}$, there is a channel $\mathcal{N}^{ij}_{T_iT_j\rightarrow ab}=\mathcal{N}^{ij}_{T_i\rightarrow a}\otimes \mathcal{N}^{ij}_{T_j\rightarrow b}$
        such that 
        \begin{equation}
            \mathcal{N}^{ij}_{T_iT_j\rightarrow ab}(\Psi_{T_iT_j})= \Psi^+_{ab}; 
        \end{equation}
        \item For each $U=\{T_i,T_j\}\in \mathcal{U}$, for all $\mathcal{N}^{ij}_{T_iT_j\rightarrow ab}=\mathcal{N}^{ij}_{T_i\rightarrow a}\otimes \mathcal{N}^{ij}_{T_j\rightarrow b}$, 
        \begin{equation}
            F(\mathcal{N}^{ij}_{T_iT_j\rightarrow ab}(\Psi_{T_iT_j}),\Psi^+_{ab}) \leq 1- \delta.
        \end{equation}
    \end{enumerate}
\end{definition}
Here, $a$ and $b$ denote output systems, produced by the local operations applied on $T_i$ and $T_j$, respectively.
We do not fix the dimensions of $a$ and $b$ in this definition, since they may vary in different scenarios.
The double index on the channels $\mathcal{N}^{ij}$ indicates that each party is aware of the identity of its partner in the pairing, so their local operations can depend on the specification of the subset held by the other party.

Given the above definition, there are various requirements we can place on $\delta$ and hence various notions of security. 
We give two notions of security below. 
\begin{definition}\label{def:weaksecurity}
    An ESS construction with \textbf{weak security} is a family of quantum states $\Psi[\mathcal{S}]$, one for each possible access structure $\mathcal{S}$, such that $\Psi[\mathcal{S}]$ is a $\delta[\mathcal{S}]>0$ secure ESS. 
\end{definition}
\begin{definition}\label{def:strongsecurity}
    An ESS construction with \textbf{strong security} comprises a fixed choice of $\delta>0$ and a family of quantum states $\Psi[\mathcal{S}]$, one for each possible access structure $\mathcal{S}$, such that $\Psi[\mathcal{S}]$ is a $\delta$ secure ESS. 
\end{definition}
Note that constructions which have their unauthorized pairs be separable also achieve strong security, since the nearest separable state to $\Psi^+$ has fidelity of at most $1/d$ with $\Psi^+$ (when $\Psi^+$ has total dimension $d^2$). 
One could also consider a third security definition, perhaps appropriate in cryptographic applications, wherein there is a family of constructions parameterized by a security parameter $\lambda$, and $\delta \rightarrow \delta_0$ as $\lambda \rightarrow \infty$, where $\delta_0$ is set by $\max_{\sigma\in \text{SEP}}F(\sigma,\Psi^+)=1/d=1-\delta_0$. 
Constructions achieving separable unauthorized pairs also have cryptographic security.

There are also many possible variants of our definition, depending on how one chooses to formalize the notions of being authorized or unauthorized.
For instance, one could instead require that authorized pairs exhibit high entanglement according to a chosen entanglement measure, while unauthorized pairs are restricted to separable states.
This definition is not directly comparable to ours: our condition on authorized pairs is stronger, while our condition on unauthorized pairs is weaker.
Fortunately, several of our constructions are strong enough to satisfy the more stringent requirements as well: for threshold schemes, we explicitly construct examples where authorized pairs can generate a specific maximally entangled state, while unauthorized pairs remain separable.

As a first example, let us revisit the entanglement sharing scheme for the $((2,2,5))$ threshold access structure described in the introduction.
This scheme is realized by the state
\begin{equation}\label{eq:PsiABCDE}
    \ket{\Psi}_{ABCDE} = \sum_{k,s} \ket{k}_A\ket{k+s}_B \ket{k+2s}_C \ket{k+3s}_D \ket{k+4s}_E
\end{equation}
with local dimension $5$ and addition and multiplication over the finite field $\mathbb{F}_5$. 
We will omit the normalization for simplicity. 
To verify that this realizes a $((2,2,5))$ scheme, we first check that every pair of type $\{2,2\}$ is authorized.
Suppose, for concreteness, that Alice holds subsystems $AB$ and Bob holds $CD$.
By a local computation that can be made unitary, they can map the state to
\begin{equation}
    \ket{\Psi}_{ABCDE} \to \sum_{k,s} \ket{s}_A\ket{k}_B \ket{s}_C \ket{k}_D \ket{k+4s}_E.
\end{equation}
Then, since they both know the identity of the other subset in the pair, they know the $E$ system is the one not involved. 
This means that they can apply another local unitary to obtain
\begin{align}
    \ket{\Psi} &\to \sum_{k,s} \ket{s}_A \ket{k+4s}_B \ket{s}_C\ket{k+4s}_D \ket{k+4s}_E \nonumber \\
    &= \sum_{s} \ket{s}_A  \ket{s}_C \sum_k \ket{k+4s}_B \ket{k+4s}_D \ket{k+4s}_E \nonumber \\
    & = \ket{\Psi^+}_{AC} \ket{\text{GHZ}}_{BDE}.
\end{align}
In particular, Alice and Bob obtain a maximally entangled state between their subsystems.
An analogous procedure shows that all other pairs of type $\{2,2\}$ are authorized as well.
We emphasize that the decoding operation producing the EPR pair depends on the identity of the excluded subsystem, as seen in the second step above.

We can also verify the type $\{1,2\}$ pairs are unauthorized. 
To see this, note that in this setting, two systems will not participate in the decoding, and those two systems can be used jointly to learn $k$ and $s$, thus decohering the superposition in the state of the decoding parties and preventing recovery of entanglement. 
More explicitly, the marginal on (for example) $ABC$ is
\begin{align}\label{eq:PsiABC}
    \Psi_{ABC} = \sum_{k,s} \ketbra{k}{k}\otimes \ketbra{k+s}{k+s}\otimes \ketbra{k+2s}{k+2s}
\end{align}
which is separable across $A$ and $BC$. 
To make this more apparent, it is useful to have Bob apply a local unitary to obtain
\begin{align}
    \Psi_{ABC} = \sum_{k,s} \ketbra{k}{k}\otimes \ketbra{k}{k}\otimes \ketbra{s}{s},
\end{align}
so that Alice and Bob only share classical randomness.

\subsection{Necessary conditions}\label{sec:knownpartnernecessary}

The monotonicity property given as \cref{def:monotonicity} is a necessary condition for ESS with a known partner. 
We formalize this as a lemma. 

\knownpartnermonotonicity*

The proof is almost trivial. 
Assume $\{T_1,T_2\}$ is authorized and that $\{T_1,T_2\}\preceq\{T_3,T_4\}$. Without loss of generality, this means $T_1\subseteq T_3$ and $T_2\subseteq T_4$.
Now Alice, who holds $T_3$, can simply restrict attention to the registers in $T_1$; Bob, who holds $T_4$, does the same with $T_2$.
Thus they can recover whenever $\{T_1,T_2\}$ can, and it follows immediately that $\{T_3,T_4\}$ is also authorized.

Let us emphasize that Alice and Bob need to agree on which smaller pairing to trace down to before recovering and to include this as part of the specification of the scheme, since they cannot coordinate their actions after the state for the ESS is distributed.
To coordinate this, Alice and Bob also need to know the identity of the full pairing $\{T_3,T_4\}$, not just their respective subsystems, so this proof only applies in the known partner case.
In fact, we will see that in the unknown partner case monotonicity no longer holds.

\subsection{Threshold schemes}\label{sec:ESSthresholdconstructions}

In this section we analyze special cases of pair access structures that admit efficient constructions, in the sense of having small dimension subsystems.
We begin with $((r,r,3r-1))$ schemes, which have particularly simple realizations, and then address the more general $((a,b,a+2b-1))$ schemes where $a$ and $b$ need not coincide (we always take $a \geq b$).

\vspace{0.2cm}
\noindent \textbf{$((r,r,3r-1))$ Entanglement sharing schemes}
\vspace{0.2cm}

Our first threshold construction is inspired by an observation about the $((2,2,5))$ scheme introduced earlier. 
The state \cref{eq:PsiABCDE} may be viewed as a superposition over values in a $(2,5)$ classical secret sharing scheme, which maps
\begin{align}\label{eq:SSks}
    (k,s)\rightarrow (k,k+s,k+2s,k+3s,k+4s).
\end{align}
Because any two shares can be used to recover $s$, Alice and Bob can each produce a copy of $\ket{s}$. 
Then to recover the EPR pair they additionally need to decouple the remaining systems from $s$. 
This is possible because the particular secret sharing scheme above has the stronger property that any two shares determine not only the secret $s$ but also the randomness $k$.

Using this idea, we can construct $((r,r,3r-1))$ schemes for all $r \geq 1$.
Our construction is based on Shamir’s classical secret sharing scheme \cite{shamir1979share}, which yields an $(r,3r-1)$ threshold structure such that the local dimension of the entangled state is the same as the local dimension of each share.

In Shamir's scheme, we pick a degree $(r-1)$ polynomial\footnote{Note that there is a notational mismatch between this equation and \cref{eq:SSks}. However, the computation is essentially equivalent.}, 
\begin{align}\label{eq:defF}
    F_{\vec{a},s}(x) = a_{r-1} x^{r-1} + a_{r-2}x^{r-2} +...+ a_1 x + s.
\end{align}
We store the secret as $F_{\vec{a},s}(0)=s$, and choose the other coefficients at random.
The arithmetic is performed in the field $\mathbb{F}_p$ for $p$ any prime larger than $3r-1$. 
We then define the entangled state
\begin{align}\label{eq:ssstate}
    \sum_{\vec{a},s} \ket{F_{\vec{a},s}(1)} ...\ket{F_{\vec{a},s}(3r-1)},
\end{align}
with each subsystem having local dimension $p$.
Now suppose that Alice gets $r$ subsystems, and Bob gets another $r$ subsystems. 
For simplicity of notation we will take these to be the first and second sets of size $r$, but similar arguments work for any two subsets of size $r$. 
Then they can act locally to produce the state
\begin{align}\label{eq:ESSkk3k-1}
    \sum_{\vec{a},s} \left(\ket{s} \ket{F_{\vec{a},s}(2r+1)}...\ket{F_{\vec{a},s}(3r-1)} \right) &\otimes \left(\ket{s} \ket{F_{\vec{a},s}(2r+1)}...\ket{F_{\vec{a},s}(3r-1)} \right) \nonumber \\
    &\otimes \left(\ket{F_{\vec{a},s}(2r+1)}...\ket{F_{\vec{a},s}(3r-1)} \right). 
\end{align}
This is allowed because the polynomial $F$ evaluated on any subset of $r$ points fixes $(s,\vec{a})$, and conversely $(s,\vec{a})$ fixes the polynomial and hence the value of the polynomial on any set of points. 
Thus the maps
\begin{align}
    \ket{F_{\vec{a},s}(1)} ...\ket{F_{\vec{a},s}(r)} \rightarrow \ket{s} \ket{F_{\vec{a},s}(2r+1)}...\ket{F_{\vec{a},s}(3r-1)}, \nonumber \\
    \ket{F_{\vec{a},s}(r+1)} ...\ket{F_{\vec{a},s}(2r)} \rightarrow \ket{s} \ket{F_{\vec{a},s}(2r+1)}...\ket{F_{\vec{a},s}(3r-1)},
\end{align}
can be implemented reversibly and hence unitarily. 
Then, notice that the systems storing $s$ are in a tensor product state with the rest of the subsystems: the remaining registers hold only $(r-1)$ distinct values of $F$, and hence store no information about $s$.
Thus the systems decouple, so that Alice and Bob hold a maximally entangled state.

To see why pairings of type $\{r-1,r\}$ or smaller cannot recover an EPR pair, note that in this case the systems that are traced out contain $r$ evaluations of the secret sharing polynomial.
From these, the environment can reconstruct both $\vec{a}$ and $s$.
Consequently, the remaining systems are left in a classical mixture over the values of $F_{\vec{a},s}(i)$, yielding a separable state.
Since pairs of type $\{r-1,r\}$ are always separable, they cannot be used to recover an EPR pair, and in fact this construction achieves strong security in the sense of \cref{def:strongsecurity}.

Note that if we try to construct a threshold scheme with parameters $((r,r,m))$ with $m>3r-1$, this construction fails. 
The reason is that the environment obtains $e:=m-2r$ shares, which for $m>3r-1$ has $e> r-1$ shares, so the environment has at least $r$ shares. 
But then the environment can recover the secret stored in superposition, which decoheres the entangled state recovered by the players. 

\vspace{0.2cm}
\noindent \textbf{Connections with Reed–Solomon codes}
\vspace{0.2cm}

\noindent The states constructed above are special codewords of quantum Reed–Solomon codes (see \cref{sec:stabilizerstuff} and \cref{sec:RScodes}).
Indeed, setting $r=k=1$ in the polynomials in \cref{eq:RSpoly}, we get
\begin{equation}
    f_{c,s}(x)=c+sx.
\end{equation}
Then setting $x_i=i$ for $i=0,\cdots,n-1$, and summing the codeword \cref{eq:RSwavefunction} over $s$ (to form the $+1$ $X$ eigenstate) we obtain exactly the $((2,2,5))$ scheme constructed in \cref{eq:PsiABCDE}.

More generally, redefining $r=r'-1$, and setting $k=1$ and $n=3r'-1$ in \cref{eq:RSwavefunction}, we get 
\begin{equation}
    f_{c,s}(x)=x^{r'-1}F_{\vec{a},s}\left(1/x\right),
\end{equation}
where $F_{\vec{a},s}$ is defined in \cref{eq:defF} and $\vec{a}$ is the reverse of $c$.
Despite the notational mismatch, the codeword \cref{eq:RSwavefunction} summed over $s$ is equivalent to \cref{eq:ssstate}.

\vspace{0.2cm}
\noindent \textbf{$((a,b,a+2b-1))$ threshold schemes}
\vspace{0.2cm}

In the following, we will generalize the above perspective to construct threshold schemes for the case when the two parties may have different sizes. 
Namely, we construct an $((a,b,a+2b-1))$ entanglement sharing scheme for arbitrary integers $a$ and $b$. 
We assume $a\geq b\geq 2$.

Consider the Reed–Solomon code with parameters $n=a+2b-1$, $k=1$ and $r = b-1$. 
Since $k=1$, we have only one $X$-type logical operator and one $Z$-type logical operator up to equivalence (see \cref{sec:RScodes}).
We pick the code state $|\Psi \rangle$ that is  stabilized by the $X$-type logical operator.
\begin{theorem}
    For $a\geq b\geq 2$, the above state $\ket{\Psi}$, defined over qudit subsystems with prime dimension $p\geq n$, is a $((a,b,a+2b-1))$ ESS threshold scheme with strong security.
\end{theorem}
\begin{proof}\,
Let us consider two arbitrary sets of parties $T_1$ and $T_2$ such that $|T_1|=a$, $|T_2|=b$ and $T_1 \cap T_2 = \emptyset$. 
Since $|T_1 \cup T_2| = a + b = n-r$, there exists an $X$-type logical operator supported on $T_1 \cup T_2$ due to \cref{lemma:RScode} (set $k=1$).
Furthermore, we can express $\bar{X}$ in terms of its tensor components on $T_1$, $T_2$ and the rest of the $n$ qudits, collected into the set denoted by $R$: 
\[ \bar{X} = \bar{X}_{T_1} \otimes \bar{X}_{T_2} \otimes I_R,\] where each $\bar{X}_{T_i}$ denotes a Pauli operator supported on $T_i$.
We have chosen $|\Psi \rangle$ to be the $+1$ eigenstate of $\bar{X}$ within the Reed–Solomon code, so   
\begin{eqnarray}
|\Psi \rangle = \bar{X} |\Psi \rangle = \bar{X}_{T_1} \otimes \bar{X}_{T_2} \otimes I_R \;|\Psi \rangle .
\end{eqnarray}
Furthermore, $|T_1|, \ |T_2| \geq r+1$, so each of $T_1$ and $T_2$ supports a $Z$-type logical operator from \cref{lemma:RScode}. Let these $Z$-type logical operators be $\bar{Z}_{T_1}, \bar{Z}_{T_2}$.
%
Since $k=1$, there is only one logical qudit. 
Therefore, $\bar{Z}_{T_1}, \bar{Z}_{T_2}$ each act on the code space as $\bar{Z}^{a_1}$ and $\bar{Z}^{a_2}$ for non-zero $a_1,a_2\in \mathbb{F}_p$.
Then define $\bar{Z}_{T_1}'
= \bar{Z}_{T_1}^{1\!/\!a_1}$ and $\bar{Z}_{T_2}'
= \left( \bar{Z}_{T_2}^{1\!/\!a_2} \right)^{\dagger}$, 
so $\bar{Z}_{T_1}'$ acts as $\bar{Z}$ on the code space, and $\bar{Z}_{T_2}'$ acts as $\bar{Z}^{\dagger}$. 
Then we have that $\bar{Z}_{T_1}' \otimes \bar{Z}_{T_2}'$ is a stabilizer,
\begin{eqnarray}
|\Psi \rangle = \bar{Z}_{T_1}' \otimes \bar{Z}_{T_2}'\otimes I_R |\Psi \rangle. 
\end{eqnarray} 
Note the asymmetry that $\bar{X}$ is supported jointly on $T_1$ and $T_2$, while $\bar{Z}$ is supported on each of $T_1$ and $T_2$. 

Since $\bar{Z}_{T_1}'$ is a logical-$Z$ operator, we have $\bar{Z}_{T_1}'\bar{X}=\omega\bar{X}\bar{Z}_{T_1}'$.
Therefore, 
\begin{equation}
    \bar{Z}_{T_1}'\bar{X}_{T_1}=\omega\bar{X}_{T_1}\bar{Z}_{T_1}'.
\label{eq:mescom1}
\end{equation}
Similarly, 
\begin{equation}
\bar{Z}_{T_2}' \bar{X}_{T_2}=\omega^{-1}\bar{X}_{T_2}\bar{Z}_{T_2}'.
\label{eq:mescom2}
\end{equation}
By \cref{lemma:stabdistillcondition}, we know that the above two conditions imply a maximally entangled state can be distilled by local operations from $T_1$ and $T_2$.

Let us now consider the case when $|T_1|+|T_2|< a+b$ and show that $\{T_1, T_2\}$ is unauthorized. 
From \cref{lemma:RScode}, they cannot support any $X$-type logical operators or stabilizer operators.
Recall from \cref{sec:stabilizerstuff} that, when reducing the state $\ket{\Psi}$ to $T_1\cup T_2$, the reduced state can be written as
\begin{align}
    \rho_{T_1T_2} = \frac{1}{d}\sum_{g\in S, \,\supp(g)\subseteq T_1\cup T_2} g,
\end{align}
where $d$ is the dimension of the $T_1\cup T_2$ Hilbert space.
In particular, only $Z$-type operators appear in the sum.
We can simultaneously diagonalize all the terms by working in the $Z$-basis.
Due to positivity, $\rho_{T_1T_2}$ is therefore a summation of product states in the $Z$ basis, which is separable.
Consequently, this scheme achieves strong security. 
\end{proof}

\subsection{General access structures for stabilizer states}\label{sec-stabiffknown}

Throughout this section, we will assume that the distributed state is a stabilizer state over $p$-dimensional qudits where $p$ is a prime number. 
Accordingly, we demand that a $p$-dimensional maximally entangled state be shared by the authorized pairs from the entanglement sharing scheme.
Following \cref{lemma:stabdistillcondition}, given the description of the distributed stabilizer state, we can determine if a disjoint pair $\{T_1,T_2\}$ is authorized.
Yet we are interested in a slightly different and more complex question: given authorized and unauthorized sets $\A$ and $\U$, is there an ESS realizing the access structure?

While we have shown in \cref{lemma:ESSmonotonicity} 
that monotonicity is necessary for an ESS, it turns out monotonicity is \emph{not} always sufficient to guarantee a stabilizer ESS.
In the following we present conditions that fully characterize the access structures which can be realized in stabilizer constructions.

We begin with the following notion. 
\begin{definition}
An authorized pair $\{T_1,T_2\}\in \mathcal{A}$ is said to be \textbf{minimal} if there does not exist another pair $\{T_1',T_2'\}\in \mathcal{A}$ such that $\{T_1',T_2'\} \prec \{T_1,T_2\}$.
\end{definition}
We denote the set of minimal authorized pairs as $\A_{\min}$.

For any minimal pair $A=\{T_1,T_2\}\in \A_{\min}$, we construct two matrices $M(A)$ and $\tilde{M}(A)$ as follows.
First, we consider $\U(A)$, a subset of the unauthorized sets $\U$, defined as:
\begin{equation}
    \U(A)=\{
    \{T_1',T_2'\}\in\U \,|\, T_1'\cup T_2'=T_1\cup T_2
    \}.
\end{equation}
The set $\U(A)$ consists of all those unauthorized pairs which are formed by repartitioning the authorized pair $A$.
The matrices $M(A)$ and $\tilde{M}(A)$ record how the unauthorized sets in $\U(A)$ are related to $A$ by repartitioning.

Denote $K_A=|\U(A)|$, $n_A=|T_1|+|T_2|$.
The matrix $M(A)$ will be a $(K_A+1)\times n_A$ matrix, whose rows are indexed by elements of $\U(A)$ and columns are indexed by elements of $T_1\cup T_2$.
The matrix $M(A)$ is defined as follows:
\begin{itemize}
    \item For each pair $U=\{T_1',T_2'\}\in\U(A)$, pick an arbitrary ordering $(T_1',T_2')$;
    \item For the first $K_A$ rows of $M(A)$: $M(A)_{U,s}=1$ if and only if party $s$ is contained in the first component of pair $U$ ($s \in T_1'$);
    $M(A)_{U,s}=0$ otherwise.
    \item The last row of $M(A)$ is all 1's.
\end{itemize}
The matrix $\tilde{M}(A)$ will be a $(K_A+2)\times n_A$ matrix, whose first $K_A+1$ rows are the same as $M(A)$, and the last row corresponds to the pair $A$ itself: $\tilde{M}(A)_{A,s}=1$ if party $s$ is contained in $T_1$, the first component of pair $A$.

The next lemma expresses that for $A$ to be authorized even while all of the repartitions of $A$ in $\U(A)$ are unauthorized, the last row of $\tilde{M}(A)$ (capturing the partition corresponding to $A$) needs to be linearly independent from the earlier rows capturing the partitions of $\U(A)$ along with the all $1$'s row (which is related to enforcing that generators of the stabilizer group commute). 

\begin{lemma}\label{lemma:rank}
    Suppose that the access pair structure $(\mathcal{A}, \mathcal{U})$ can be realized using a stabilizer construction. Then, $\forall A\in\A_{\min}$, $\rank M(A)<\rank \tilde{M}(A)$.
    Here the rank is defined over $\mathbb{F}_p$.
\end{lemma}

\begin{proof}\,
Consider an arbitrary authorized pair $A \in\A_{\min}$, and order $A$ as $(T_1,T_2)$. 
(1) By \cref{remark:tripartitestabilizer}, there exists a unitary $U = U_{1} \otimes U_2 \otimes U_3$ acting locally on $T_1 \otimes T_2 \otimes R$ where $R$ includes all the systems not in $T_1 \cup T_2$, that transforms the original ESS state to one including a maximally entangled state in $T_1$ and $T_2$. 
Let $T_1 = T_{11} T_{12}$ and $T_2 = T_{21} T_{22}$ and the output maximally entangled state live on $T_{11} T_{21}$. 
(2) Recall that in the stabilizer formalism, if we transform the state $\ket{\psi}$ to $U\ket{\psi}$, each stabilizer is transformed from $S$ to $USU^\dagger$. Combining the above two facts (1) and (2), and using the stabilizers for the maximally entangled state, there are two stabilizer generators $V,W$ for the initial ESS state such that 
\begin{align*}
UWU^\dagger & = X_{T_{11}} \otimes I_{T_{12}} \otimes X_{T_{21}} \otimes I_{T_{22}} \otimes I_R, \\
UVU^\dagger & = Z_{T_{11}} \otimes I_{T_{12}} \otimes Z^{\dagger}_{T_{21}} \otimes I_{T_{22}} \otimes I_R \,.
\end{align*}
Inverting the conjugation by $U$ and using that $U = U_{1} \otimes U_2 \otimes U_3$, we obtain 
\begin{align*}
W & = \left[ U_1^\dagger (X_{T_{11}} \otimes I_{T_{12}})\, U_1 \right]_{T_1} \otimes \left[ U_2^\dagger (X_{T_{21}} \otimes I_{T_{22}})\, U_2 \right]_{T_2} \otimes I_R, \\
V & = \left[ U_1^\dagger (Z_{T_{11}} \otimes I_{T_{12}})\, U_1 \right]_{T_1} \otimes \left[ U_2^\dagger (Z^{\dagger}_{T_{21}} \otimes I_{T_{22}})\, U_2\right]_{T_2}  \otimes I_R \,.
\end{align*} 
Defining 
\begin{align*} 
W_{T_1} &= U_1^\dagger (X_{T_{11}} \otimes I_{T_{12}})\, U_1 \,,~~~ 
W_{T_2} = U_2^\dagger (X_{T_{21}} \otimes I_{T_{22}})\, U_2, \\ 
V_{T_1} &= U_1^\dagger (Z_{T_{11}} \otimes I_{T_{12}})\, U_1 \,,~~~ 
V_{T_2} = U_2^\dagger (Z^{\dagger}_{T_{21}} \otimes I_{T_{22}})\, U_2 \,,
\end{align*}
we have
\begin{align}
    V_{T_1}W_{T_1}=\omega W_{T_1}V_{T_1}&,~~
    V_{T_2}W_{T_2}=\omega^{-1} W_{T_2}V_{T_2}.\label{eq:rank1}
\end{align}
Define a vector $x \in \mathbb{F}_p^{n_A}$, indexed by elements of $T_1\cup T_2$, such that the $s$-th entry $x_s$ is set by 
\begin{equation}
    \omega^{x_s}=V_{s}W_{s}V_{s}^{-1}W_{s}^{-1},
\end{equation}
where $V_s$ ($W_s$) is the tensor component of $V$ ($W$) on party $s$.
The vector $x$ records the component-wise commutation properties of $V$ and $W$. 
The fact that $V$ and $W$ commute, $VW=WV$, becomes
\begin{align}
    z_0x=0
\end{align}
where $z_0$ is the second to last row of $\tilde{M}(A)$ (the all $1$'s vector). 
Meanwhile the fact that the $T_1$ components on their own pick up a phase of $\omega^1$ becomes 
\begin{equation}
    z_{A}x=1,
\end{equation}
where $z_A$ is the last row of $\tilde{M}(A)$.
  
To finish our proof, we take the $V,W$ defined above, and consider each $U=\{T_1',T_2'\}\in\U(A)$. As before, we express $V$ and $W$ as a product operator, but now according to the tripartition $(T_1', T_2',R)$:
\[ 
V = V_{T'_1} \otimes V_{T'_2} \otimes I_R \,, ~~~
W = W_{T'_1} \otimes W_{T'_2} \otimes I_R \,.
\]
    In contrast to \cref{eq:rank1}, we claim we now have 
    \begin{align}
        V_{T_1'}W_{T_1'}=W_{T_1'}V_{T_1'}&,~~
        V_{T_2'}W_{T_2'}=W_{T_2'}V_{T_2'}.\label{eq:rank2}
    \end{align}
We can prove the above by contradiction. Suppose $V_{T_1'}W_{T_1'}= \omega^r W_{T_1'}V_{T_1'}$ for some $r \not\equiv 0 \bmod p$. 
Then, because $V,W$ commute (both are stabilizers of the initial state) we have $V_{T_2'}W_{T_2'}= \omega^{-r} W_{T_2'}V_{T_2'}$. 
By \cref{lemma:stabdistillcondition} this implies there is a distillable EPR pair between $T_1'$ and $T_2'$.

The first equation in \cref{eq:rank2} can be represented as
\begin{equation}\label{eq:rank1.1}
    z_Ux=0,
\end{equation}
where $z_U$ is the row of index $U$ in the matrix $M(A)$.
The second equation is redundant, since $VW=WV$, or equivalently, $z_0x=0$.
Therefore, there exists an $x$ such that $M(A)x=0$ but $\tilde{M}(A)x\neq 0$.
This implies $\rank M(A)<\rank \tilde{M}(A)$.
\end{proof}

For intuition, let us consider a simple example of an access structure ruled out by this lemma. 
Suppose we have three parties $S_1,S_2,S_3$ and wish to construct a scheme with only one authorized pair $A=\{\{S_1\},\{S_2,S_3\}\}$, while $\{\{S_2\},\{S_1,S_3\}\}$ and $\{\{S_3\},\{S_1,S_2\}\}$ are unauthorized. 
Monotonicity holds, but the existence of such an ESS contradicts \cref{lemma:rank}. 
For the proposed access structure, we would have
\begin{equation}
    M(A)=\begin{pmatrix} 0&1&0 \\ 0&0&1\\1&1&1 \end{pmatrix},
    ~~
    \tilde{M}(A)=\begin{pmatrix}0&1&0 \\ 0&0&1\\ 1&1&1 \\1&0&0 \end{pmatrix}.
\end{equation}
Hence $\rank M(A)=\rank\tilde{M}(A)=3$, so the lemma shows this access structure cannot be realized by any stabilizer state. 

Next we show that monotonicity along with \cref{lemma:rank} are sufficient in the stabilizer setting.
\begin{theorem}\label{thm-stabiffknown}
    An access structure $(\mathcal{A},\mathcal{U})$ can be realized by a stabilizer entanglement sharing scheme with strong security and known partner if and only if $(\mathcal{A}, \U)$ satisfies 
    \begin{enumerate}
        \item \textbf{Monotonicity:} For all pairs $A'$ such that there exists $A\in \mathcal{A}$ with $A\preceq A'$ we have that $A'\in \mathcal{A}$. 
        \item \textbf{Rank condition:} $\forall A\in\A_{\min}$ we have $\rank M(A)<\rank \tilde{M}(A)$ (i.e. the condition of \cref{lemma:rank}).
    \end{enumerate}
\end{theorem}
\begin{proof}\,
    By \cref{lemma:ESSmonotonicity} and \cref{lemma:rank}, both conditions are necessary. 
    In the following, we prove the sufficiency by  
    constructing an explicit entanglement sharing scheme with access structure $(\mathcal{A}, \U)$. 
    Since $\A$ satisfies monotonicity, $\A_{\min}$ is well-defined. 
    The construction consists of a tensor product of many mixed stabilizer states, each of which authorizes a single minimal pair in $\mathcal{A}$ (and those implied by monotonicity) and no pair in $\U$.  
    
    For each $A=\{T_1,T_2\}\in\A_{\min}$, we construct $M(A)$ and $\tilde{M}(A)$ as defined before.
    \Cref{lemma:rank} implies that there exists $x$ such that
    \begin{align}\label{eq:rank3}
    z_0x=0,& \,\,\,\,\,\,\,z_Ax\neq 0, \nonumber \\
    \forall \, U\in\U(A),& \,\,\, ~~z_Ux=0.
    \end{align}
    Now, we assign each party a qudit of dimension $p$, and define Pauli operators $V$ and $W$ as
    \begin{equation}\label{eq:VAWA}
        V^A=\otimes_{s\in T_1\cup T_2} Z,~~
        W^A=\otimes_{s\in T_1\cup T_2} X^{x_s}.
    \end{equation}
    \Cref{eq:rank3} implies that for $\ell \equiv z_Ax\neq 0$ we have
    \begin{align}
        &V_{T_1}^AW_{T_1}^A=\omega^{\ell} W^A_{T_1}V^A_{T_1},~~
        V^A_{T_2}W^A_{T_2}=\omega^{-\ell} W^A_{T_2}V^A_{T_2};\label{eq:commcond1}\\
        &\forall \{T_1',T_2'\}\in \U(A), \,\,\,V^A_{T_1'}W^A_{T_1'}=W^A_{T_1'}V^A_{T_1'},~~
        V^A_{T_2'}W^A_{T_2'}=W^A_{T_2'}V^A_{T_2'}~. \label{eq:commcond2}
    \end{align}    
    Let $\rho_A$ be the mixed stabilizer state with stabilizer generated by $V^A$ and $W^A$ only. 
    According to \cref{eq:mixedstabilizerstate2}, $\rho_A \propto \left(\sum_{i=0}^{p-1}(V^A)^i\right)\left(\sum_{j=0}^{p-1}(W^A)^j\right)$. 
    Note also that the systems outside of $A$ are in the maximally mixed state.  
    If $\rho_A$ is distributed to the $n$ parties, the following holds:
    \begin{itemize}
        \item $A$ is authorized. This is due to \cref{eq:commcond1} and the sufficient condition for decoding entanglement given as \cref{lemma:stabdistillcondition}.
        \item $\forall U\in \U(A)$, $U$ is unauthorized. This follows because any two stabilizers of $\rho_A$ will have the form $S_1=(V^A)^{i_1}(W^A)^{j_1}$, $S_2=(V^A)^{i_2}(W^A)^{j_2}$, and by \cref{eq:commcond2} the restrictions of these stabilizers to $T_1'$ or $T_2'$ will always commute. Then applying the converse statement in \cref{lemma:stabdistillcondition} implies there is no recoverable EPR pair. By \cref{remark:tripartitestabilizer}, this means the state across the $T_1', T_2'$ partition is separable, as needed. 
        \item $\forall U\in \U \setminus \U(A)$, $U$ is unauthorized. 
        Suppose by contradiction that $U=\{T_1',T_2'\}$ is authorized.  Since the systems outside of $T_1\cup T_2$ contain only the maximally mixed state, if we let $T''_1 = T'_1 \cap (T_1\cup T_2)$ and $T''_2 = T'_2 \cap (T_1\cup T_2)$, $U''=\{T''_1,T''_2\}$ is also authorized. If $T''_1 \cup T''_2 = T_1\cup T_2$, $U'' \in \U(A)$ and cannot be authorized by the previous item, so, $T''_1 \cup T''_2 \subsetneqq T_1\cup T_2$. But now $V^A$, with $Z$'s in each tensor component, is not supported on $T''_1 \cup T''_2$ and the reduced state on $T''_1 \cup T''_2$ is $\propto {\Tr}_{R''} \sum_{i=0}^{p-1}(W^A)^i$ where $R''$ contains all the systems not in $T''_1 \cup T''_2$. But ${\Tr}_{R''} \sum_{i=0}^{p-1}(W^A)^i$ is separable along any bipartition, so $U''=\{T''_1,T''_2\}$ is unauthorized, a contradiction.  
    \end{itemize}
    We repeat the procedure for each $A\in\A_{\min}$, with the ESS given by the state 
    $$\bigotimes_{A\in\A_{\min}} \rho_A.$$
    In the above, each party receives a share of each $\rho_A$. 
    For each unauthorized pair $U$, the reduced state is a tensor product of separable states and it remains unauthorized. 
    For each authorized pair $A'$, we have that there exists $A$ with $A\preceq A'$ and $A \in \A_{\min}$ so $A'$ allows for decoding a maximally entangled state using just $\rho_A$.
 \end{proof}

\section{Entanglement sharing schemes with an unknown partner}\label{sec:ESSunknown}

\subsection{Definition and first example}\label{sec:ESSdef}

In this section we discuss entanglement sharing schemes in the setting where the partner is unknown. 
In other words, a pairing $\{T_1,T_2\}$ now means that Alice obtains $T_1$ but does not know $T_2$. 
Similarly, Bob obtains $T_2$ but does not know $T_1$.
We state this more carefully below. 

\begin{definition}
    We say a state $\Psi_{S_1...S_n}$ is a $\delta$-secure \textbf{entanglement sharing scheme (ESS) with an unknown partner} and with access structure $\mathcal{S}$ if the following hold. 
    \begin{enumerate}
        \item There exists a family of channels $\{\mathcal{N}_{T_i\rightarrow X_i}\}$ such that for each $A_{k}=\{T_i,T_j\}\in \mathcal{A}$, 
        \begin{align}
            \mathcal{N}_{T_iT_j\rightarrow X_iX_j}(\Psi_{T_iT_j})= \Psi^+_{X_iX_j}
        \end{align}
        where we define $\mathcal{N}_{T_iT_j\rightarrow X_iX_j}=\mathcal{N}_{T_i\rightarrow X_i}\otimes \mathcal{N}_{T_j\rightarrow X_j}$.
        \item For each $U_k=\{T_i,T_j\}\in \mathcal{U}$, and all possible channels $\mathcal{M}^{k}_{T_iT_j\rightarrow X_iX_j} = \mathcal{M}^k_{T_i\rightarrow X_i}\otimes \mathcal{M}^k_{T_j\rightarrow X_j}$ we have
        \begin{align}
            F(\mathcal{M}^{k}_{T_iT_j\rightarrow X_iX_j}(\Psi_{T_iT_j}),\Psi^+_{X_iX_j}) \leq 1-\delta.
        \end{align}
    \end{enumerate} 
\end{definition}
Note that the notions of strong and weak security (\cref{def:weaksecurity} and \cref{def:strongsecurity}) can be applied to the context of ESS with unknown partner. 

We can illustrate the distinction between entanglement sharing with a known and unknown partner with a simple example. 
Consider the access structure with $\{\{S_1\},\{S_2\}\}$, $\{\{S_1\},\{S_3\}\}$, and $\{\{S_1\},\{S_2,S_3\}\}$ authorized. 
This can be extended to satisfy monotonicity, and it is easy to construct for it an entanglement sharing scheme with a known partner. 
However in the context of entanglement sharing with an unknown partner the access structure is impossible to realize: Alice, who holds $S_1$, must prepare a system $X_1$ to output which must be maximally entangled with both $X_2$ and $X_3$, the systems returned by the parties holding $S_2$ and $S_3$ respectively, since she doesn't know which one she is trying to entangle with.
But the monogamy of entanglement means this is impossible, so there can be no such scheme. 

For other access structures, it is initially not clear if they can be realized as an entanglement sharing scheme with unknown partner. 
For instance consider a $((2,2,5))$ threshold access structure. 
Suppose Alice receives shares $\{S_1, S_2\}$. 
Then, Bob could receive any of three possible subsets: $\{S_3,S_4\}$, $\{S_4,S_5\}$, or $\{S_3,S_5\}$. 
Now there is no obvious violation of monogamy, and in fact we can perform entanglement sharing with an unknown partner in at least these cases by using the following strategy: take a single EPR pair $\Psi_{AB}^+$, and encode $A$ into a $((2,2))$ scheme and $B$ into a $((2,3))$ scheme. 
Then give the shares $A$ is encoded into to $S_1$ and $S_2$, and the shares $B$ is encoded into to $S_3,S_4$ and $S_5$. 
While this realizes some of the authorized sets of the $((2,2,5))$ scheme, it is not clear if all of them can be realized simultaneously. 

Before addressing the general question of which access structures can be realized for ESS with an unknown partner, let us first comment on two subtleties.

First, in the context of an unknown partner, the possibility of distilling an EPR pair from a given authorized pair can depend not just on the state which has been shared, but also on which other authorized pairs are included in the access structure $\mathcal{A}$. 
In other words, given a state, the access structure may not be unique.
To illustrate this, consider a case with four parties $1,2,3,4$, and suppose we distribute the state
\begin{align}
    \ket{\Psi^+}_{12}\otimes \ket{\Psi^+}_{34}.
\end{align}
For this state, we can have an authorized pair $\{\{1,3\},\{2\}\}$ or an authorized pair $\{\{1,3\},\{4\}\}$, but we cannot include both in the pair access structure simultaneously. 
If we do, someone holding $\{1,3\}$ will be unsure if they should output qubit $1$ or qubit $3$ to prepare entanglement with the partner subsystem. 
This subtlety means in particular that access structures for unknown partner ESS may not be monotone. 
For instance, suppose we have authorized pairs $\{\{1\},\{2\}\}$, $\{\{3\},\{4\}\} \in \mathcal{A}$. 
By monotonicity we would be led to conclude that both  $\{\{1,3\},\{2\}\}$ and $\{\{1,3\},\{4\}\}$ should be authorized as well, but in fact we can only have one of these two in $\mathcal{A}$ given the above state. 

Second, in the definition of ESS with unknown partners, we have restricted the output to be the standard Bell state.
As an apparently more general definition, we could demand only that they output any maximally entangled state.
However, we will show in \cref{lem:calibration} that these two definitions are in fact equivalent.

\subsection{Necessary conditions}\label{sec:neces-ESS-unknown}

In this section we develop an understanding of which pair access structures can be realized as entanglement sharing schemes with an unknown partner. 
We start by proving necessary conditions on these access structures.

There are two classes of necessary conditions.
First, there are conditions which involve only the pattern of authorized sets.
For example, we argued above that $\{\{S_1\},\{S_2\}\}$ and $\{\{S_1\},\{S_3\}\}$ cannot both be authorized if $S_2\cap S_3=\emptyset$, due to monogamy of entanglement. 

Second, there are conditions that appear when we focus on access structures that are maximal, i.e., that cannot be extended further. 
More precisely, we give the following definition. 
\begin{definition}\label{def:maximality}
    A set of authorized pairs $\A$ is \textbf{maximal} with respect to a state $\rho$ and a valid family of channels $\{\N_T\}$, if there does not exist a proper superset $\A'\supset \A$ and a family of channels $\{\N_T\,|\,T\in\cup \A'\}$ that extends (or equals) $\{\N_T\}$ such that $\A'$ and $\{\N_T\,|\,T\in\cup \A'\}$ define an ESS with unknown partner for $\rho$. 
\end{definition} 
An analogous maximality condition already appears in quantum secret sharing, which leads to the monotonicity condition. 
For ESS with unknown partner, we noted above that monotonicity fails. 
Instead, maximality leads to more subtle conditions.

Before proving our necessary conditions, we give the following simple statements. 
\begin{lemma}\label{lem:transpose}
Let $d\geq 2$. 
\begin{enumerate}
    \item There does not exist a $U\in SU(d)$ and a nonzero $\ket{\psi}\in\mathbb{C}^d$ such that $VU\ket{\psi}=UV^{T}\ket{\psi}, ~\forall\, V\in SU(d)$.
    \item There does not exist a nonzero $\ket{\psi}\in\mathbb{C}^d\otimes \mathbb{C}^d$ such that $(V\otimes I)\ket{\psi}=(I \otimes V)\ket{\psi}, ~\forall\, V\in SU(d)$.
    \item If $\ket{\psi}\in\mathbb{C}^d\otimes \mathbb{C}^d$ satisfies $(V\otimes I)\ket{\psi}=(I\otimes V^T)\ket{\psi},~\forall \,V\in SU(d)$, then $\ket{\psi}$ is proportional to the EPR state.
\end{enumerate}
\end{lemma}
\begin{proof}\,
    (1) We prove the claim by contradiction. So, suppose the claim is true for some $U$. We first show that the claim also holds for $U=I$. To do this, first apply the claim for $V=U^*$ so that 
    \begin{align}
    U^*U \ket{\psi} = \ket{\psi}.  \label{eq:ustaru}
    \end{align}
    Here $U^{*}$ is the complex conjugate of $U$.  
    Next, consider an arbitrary $W \in SU(d)$, and apply the claim for $V = WU^{*}$ to obtain
    \begin{align}
(WU^{*})U\ket{\psi}=U(WU^{*})^{T}\ket{\psi}=UU^{\dagger}W^{T}\ket{\psi}=W^{T}\ket{\psi}.
    \end{align}
    But the far-left expression is $W \ket{\psi}$ due to (\ref{eq:ustaru}).  So, 
    $W^{T}\ket{\psi}=W\ket{\psi}$ for all $W$ in $SU(d)$ as claimed. Furthermore, this equation holds for all $W \in U(d)$, since we can replace $W$ by $e^{i\theta}W$.  So $\forall W \in U(d)$, 
    $W^{*}W\ket{\psi}=\ket{\psi}$.
    
    Now choose $W = \frac{X+Y}{\sqrt{2}}\oplus {I}_{d-2}$, which is indeed a unitary. 
    We have $W^{*}W=(iZ)\oplus {I}_{d-2}$.
    Therefore, the first two components of $\ket{\psi}$ must be zero.  We can repeat the argument for the $i$-th and $(i+1)$-th entries for $i=2,\cdots,d-1$ and 
    conclude that $\ket{\psi}=0$, which is a contradiction. 

    (2) Under the isomorphism between $\mathbb{C}^d\otimes \mathbb{C}^d$ and $M(d,\mathbb{C})$, the claim is equivalent to the nonexistence of (non-zero) $A\in M(d,\mathbb{C})$ such that $VA=AV^T$ for all $V\in SU(d)$.
    To prove the claim, we suppose $VA=AV^T$ for all $V\in SU(d)$ and prove that $A=0$. 
    Note first that the matrices in $SU(d)$ linearly span $M(d,\mathbb{C})$, so by linearity we have also that $MA=AM^T$ for all $M\in M(d,\mathbb{C})$. 
    But then choose $M=\ketbra{i}{j}$.   
    Then for $A=\sum_{k,\ell} A_{k,\ell} \ketbra{k}{\ell}$, we obtain
    \begin{align}
        \sum_{\ell} A_{j,\ell} \ketbra{i}{\ell} = \sum_{k} A_{k,j} \ketbra{k}{i}.
    \end{align}
    But now multiply on the left by $\ketbra{x}{y}$ with $y\neq i$, giving
    \begin{align}
        0 = A_{y,j}
    \end{align}
    Repeating this for all possible $y,j$ we obtain $A=0$. 

    (3) Let $A$ be the matrix corresponding to $\ket{\psi}$ under the isomorphism between $\mathbb{C}^d\otimes \mathbb{C}^d$ and $M(d,\mathbb{C})$.
    We have that $AV=VA$ for all $V\in SU(d)$.
    Since $SU(d)$ linearly spans the whole $M(d,\mathbb{C})$, we conclude that $A$ is in the center of $M(d,\mathbb{C})$, which contains only the matrices $\alpha I$ for $\alpha$ scalar.
    Taking the isomorphism back to $\mathbb{C}^d\otimes \mathbb{C}^d$, this means $\ket\psi\propto\ket{\Psi^+}$.
\end{proof}

For necessary conditions of the first type (not using maximality), we have the following \textbf{no odd cycles} and \textbf{monogamy} conditions.
\begin{lemma}[No odd cycles]\label{lemma:nooddcyclesESS}
    Consider the graph $G_{\mathcal{A}}$ describing the pair access structure of an entanglement sharing scheme with unknown partner. Then, for $\mathcal{S}=(\mathcal{A},\mathcal{U})$ to be a valid pair access structure, $G_{\mathcal{A}}$ must not contain an odd length cycle. 
\end{lemma}

\begin{proof}\,
We have outlined the intuitions behind the proof in \cref{sec:firstexample}.  
We now provide a detailed proof by contradiction.
Suppose $G_\A$ contains a cycle: $\{T_1,T_2\},\dotsb,\{T_{k-1},T_k\}, \{T_{k},T_1\}$ for some odd length $k$.
We may dilate each distillation channel into a unitary $U_{T_i}$ and purify the shared state to $|\Psi \rangle$ (by adding a party that is never in any authorized set), so that a maximally entangled state can be distilled: 
\begin{equation}\label{eq:distilledepr}
U_{T_i} U_{T_{i+1}} |\Psi \rangle = |\Psi^+ \rangle_{\bar{T}_i \bar{T}_{i+1}} |\Psi' \rangle \ ,
\end{equation}
where $\bar{T}_i$ denotes a $d_E$-dimensional subsystem of $T_i$ and $d_E$ is the dimension of the output entangled state. 

Using the transpose trick on the EPR state, $V_{\bar{T}_i}  | \Psi^+ \rangle_{\bar{T}_i \bar{T}_{i+1}} = V^T_{\bar{T}_{i+1}} |\Psi^+ \rangle_{\bar{T}_i \bar{T}_{i+1}} $ for arbitrary $V \in SU(d_E)$. 
Thus by applying $V$ on the distilled EPR state in \cref{eq:distilledepr}, we obtain
\begin{equation}
V_{\bar{T}_i} U_{T_i} U_{T_{i+1}} |\Psi \rangle 
= V^T_{\bar{T}_{i+1}} U_{T_i} U_{T_{i+1}} | \Psi \rangle.
\end{equation}
Left-multiplying $U_{T_i}^\dagger  U_{T_{i+1}}^\dagger$ on both sides and using $T_i \cap T_{i+1}=\emptyset$, we get:
\begin{equation}\label{eq:transpose}
U_{T_i}^\dagger V_{\bar{T}_i} U_{T_i} |\Psi \rangle = U^\dagger_{T_{i+1}} V^T_{\bar{T}_{i+1}} U_{T_{i+1}}
|\Psi \rangle.
\end{equation}
Namely, each edge in the path flips a $V$ on one end to $V^T$ on the other end. 

By repeating the process $k$ times along the closed walk $T_1,T_2,T_3 \dotsb T_{k},T_1$, 
we obtain  
\begin{equation}
U^\dagger_{T_1} V_{\bar{T}_1} U_{T_1} |\Psi \rangle
= U^\dagger_{T_2} V_{\bar{T}_2}^T U_{T_2} |\Psi \rangle
= \dotsb
= U^\dagger_{T_1} V_{\bar{T}_1}^T U_{T_1} |\Psi \rangle\ .
\end{equation}
This implies that there exists a state $|\Psi' \rangle = U_{T_1} |\Psi \rangle$ such that $V_{\bar{T}_1} |\Psi' \rangle = V_{\bar{T}_1}^T|\Psi' \rangle$ for arbitrary unitary $V_{\bar{T}_1}$. 
Schmidt decomposing $\ket{\Psi'}$ across $\bar{T}_1$ and the rest of the Hilbert space, we see that there must exist a (non-zero) vector $\ket{\psi}_{\bar{T}_1}$ such that $V_{\bar{T}_1} \ket{\psi}_{\bar{T}_1} = V_{\bar{T}_1}^T\ket{\psi}_{\bar{T}_1}$ for all $V\in SU(d_E)$.
This leads to a contradiction, due to \cref{lem:transpose}(1).
\end{proof}

\begin{remark}\label{remark:generalizednocycles}
    The above lemma continues to hold even if we only require that the recovery operations return, for each authorized pair in the cycle, any fixed maximally entangled state. 
    This follows by writing an arbitrary maximally entangled state $\ket{\Phi}$ as $\ket{\Phi}=\mathcal{I}\otimes W\ket{\Psi^+}$ and then applying the transpose trick again for each edge in the cycle. After going around the cycle, the operator now returns as $A \mathcal{O}^T A^\dagger$. But a slight generalization of lemma \ref{lem:transpose} shows that for arbitrary $\mathcal{O}$, there can be no state $\ket{\Psi}$ such that $A \mathcal{O}^T A^\dagger\ket{\Psi}=\mathcal{O}\ket{\Psi}$, a contradiction. 
\end{remark}

\begin{lemma}[Monogamy]\label{lemma:monogamyESSWKP}
    If there is an even length path $\{T_1,T_2\}$,...,$\{T_{k-1},T_{k}\}\in G_{\mathcal{A}}$ ($k$ odd), for $\A$ the authorized pairs in an unknown partner ESS, then $T_1\cap T_{k}\neq\emptyset$.
\end{lemma}
\begin{proof}\,
    Similar to the proof of the no odd cycle condition, the hypothesis implies that there are 
    unitaries $U_{T_i}$ acting on $T_i$ for distilling   entanglement, so that  
    \cref{eq:transpose} holds for all $i\in \{1,2,\cdots, k-1\}$ and $V\in SU(d_E)$.

    Now, since the length of path is even, we have
    \begin{align}
        U^\dagger_{T_1} V_{\bar{T}_1} U_{T_1} \ket{\Psi}=U^\dagger_{T_k} V_{\bar{T}_k} U_{T_k}\ket{\Psi} 
        \,.
    \end{align}
    We now assume by contradiction that $T_1\cap T_{k}=\emptyset$. Then we can multiply both sides of the above by $U_{T_k} U_{T_1}$ and commute the operators around to obtain 
    \begin{align} \label{eq:monotransit}
        V_{\bar{T}_1}\ket{\Phi}=V_{\bar{T}_k}\ket{\Phi}
        \,,
    \end{align}
    where $\ket{\Phi}=U_{T_k}U_{T_1}\ket{\Psi}$.
    Now use the Schmidt decomposition across $\bar{T}_1\bar{T}_k$ and the rest of the Hilbert space to conclude there is a non-zero vector $\ket{\psi}_{\bar{T}_1\bar{T}_k}$ such that $V_{\bar{T}_1}\ket{\psi}_{\bar{T}_1\bar{T}_k}=V_{\bar{T}_k}\ket{\psi}_{\bar{T}_1\bar{T}_k}$.  Recall that this holds for all $V\in SU(d_E)$, 
    so we contradict \cref{lem:transpose}(2). 
\end{proof}

We make two remarks before stating the next lemma.  

{\bf Remark 1.} Recall that in our definition of an access structure, we imposed that all pairings $\{T_i,T_j\}$ in $\A$ or $\U$ have $T_i\cap T_j =\emptyset$.
Under this requirement, monogamy turns out to imply the absence of odd cycles. 
Indeed, suppose there exists an odd-length cycle, $\{T_1, T_2\}, \ldots, \{T_k, T_1\}$ (with $k$ odd).
Then, by the monogamy condition applied to the edges $\{T_1, T_2\}, \ldots, \{T_{k-1}, T_k\}$, we must have $T_1 \cap T_k \neq \emptyset$.
However, since $\{T_k, T_1\} \in \mathcal{A}$, the definition of the access structure implies $T_1 \cap T_k = \emptyset$, leading to a contradiction.
Nonetheless, the absence of odd cycles remains a valid constraint under certain more general definitions of access structures (for example, where overlaps are allowed as long as the distillation channels commute and do not act on the outputs of other channels).
Therefore, we will retain both conditions in our discussion.

{\bf Remark 2.} If $k$ is even and $T_1\cap T_{k}=\emptyset$, a similar argument as the proof for \cref{lemma:monogamyESSWKP} goes through with $V_{\bar{T}_k}$ replaced by $V^T_{\bar{T}_k}$.  
Looking at part (3) of \cref{lem:transpose}, this suggests $\{T_1,T_{k}\}$ is authorized. 
This is proved in the following necessary condition.

\begin{lemma}[Transitivity]\label{lemma:transitiveESS}
    Consider a set of authorized pairs $\A$ in an unknown partner ESS.
    If there is a path of edges $\{T_1,T_2\}$,...,$\{T_{k-1},T_{k}\}\in G_{\mathcal{A}}$ with odd length and $T_1\cap T_{k}=\emptyset$, then $\{T_1,T_{k}\}$ is authorized.\footnote{We note that, due to \cref{lemma:monogamyESSWKP}, the condition of length being odd is in fact redundant given $T_1\cap T_{k}=\emptyset$. 
However, related to the comment below \cref{lemma:monogamyESSWKP}, we also retain both conditions in the statement of the current lemma.} 
\end{lemma}

\begin{proof}\,
    Following {\bf Remark 2} above, for an arbitrary $V \in SU(d_E)$, we obtain analogously to  \cref{eq:monotransit}, 
    \begin{align} \label{eq:monotransitT}
        V_{\bar{T}_1}\ket{\Phi}=V^T_{\bar{T}_k}\ket{\Phi}
        \,,
    \end{align}
    We apply the Schmidt decomposition across $\bar{T}_1\bar{T}_k$ and the rest of the Hilbert space ($R$) to write
    \begin{align} \label{eq:Schdecomp}
        \ket{\Phi}_{\bar{T}_1\bar{T}_kR}=\sum_i \lambda_i \ket{\psi_i}_{\bar{T}_1\bar{T}_k}\ket{\phi_i}_R \,.  
    \end{align}
    Using \cref{eq:monotransitT} and the orthogonality of 
    the $\ket{\phi_i}_R$'s, for every $i$, $V_{\bar{T}_1}\ket{\psi_i}_{\bar{T}_1\bar{T}_k}= V^T_{\bar{T}_k} \ket{\psi_i}_{\bar{T}_1\bar{T}_k}$. 
    \Cref{lem:transpose}(3) then implies that $\ket{\psi_i}=\ket{\Psi^+}$.  Substituting back into (\ref{eq:Schdecomp}), $\ket{\Phi}=\ket{\Psi^+}_{\bar{T}_1\bar{T}_k}\otimes \sum_i\lambda_i \ket{\phi_i}$.  
    Finally, recall that $\ket{\Phi}=U_{T_k}U_{T_1}\ket{\Psi}$, so the maximally entangled state will indeed be the output if the pair $\{T_1,T_{k}\}$ is called to decode, so, 
    $\{T_1,T_{k}\}$ is indeed authorized.
    \end{proof}

Recall from \cref{def:flatten} 
that $\cup\A$ denotes the flattening of $\A$.  The following two necessary conditions make use of maximality.
\begin{lemma}[Weak monotonicity]\label{lemma:weakmono}
    Consider a maximal set of authorized pairs $\A$ in an unknown partner ESS.
    If $\{T_1, T_2\}\in \mathcal{A}$, $T_1\subseteq T_3$ and $T_3\cap T_2=\emptyset$, then $T_3\in\cup\A$.
\end{lemma}
\begin{proof}\,
    Suppose we have constructed a scheme defined by the state $\rho$ such that $\mathcal{A}$ is maximal for $\rho$ (so only $\mathcal{A}$ is authorized, and no further pair can be).      
    Assume the contrary that $T_3 \not\in\cup\A$, that is, there is no $T_4$ such that $\{T_3,T_4\}\in\mathcal{A}$.
    Then define the distillation channel associated with $T_3$ as $\mathcal{N}_{T_3}=\mathcal{N}_{T_1}$.
    Since also $T_2\cap T_3=\emptyset$ it follows that $\mathcal{N}_{T_3}\otimes \mathcal{N}_{T_2}$ distills an EPR pair. 
    Hence we can add $\{T_3, T_2\}$ into $\mathcal{A}$.
    This contradicts the maximality of $\mathcal{A}$, so $T_3$ must have a partner and be part of an authorized pair. 
\end{proof}

As a corollary, if $T\notin\cup\A$ then $T^c\notin\cup\A$. 
To see this, suppose 
$T\notin\cup\A$ and 
$T^c\in\cup\A$. 
Then, the latter implies that there exists $T'$ such that $\{T',T^c\}\in\A$ and therefore $T'\subseteq T$. \Cref{lemma:weakmono} then implies $T\in\cup\A$ (with $T_3=T, T_1=T', T_2=T^c$), a contradiction.

For later reference, we record another consequence of maximality, which further constrains the entanglement structure of the state $\rho$.
\begin{lemma}\label{lemma:indistillable}
    Suppose $\A$ is a maximal set of authorized pairs.
    If $R\notin\cup\A$, then one cannot distill an EPR pair from $\{R, R^c\}$ on $\rho$, in the unknown partner settings.
\end{lemma}
\begin{proof}\,
    Otherwise, we can always add the pair $\{R, R^c\}$ to $\A$ (since $R\notin\cup\A$ and $R^c\notin\cup\A$, there is no confusion), violating maximality.
\end{proof}

Finally, we return to the subtlety mentioned at the end of \cref{sec:ESSdef}.
The following lemma shows that restricting the distillation outputs to be the standard Bell state, rather than any maximally entangled state, does not cause loss of generality.
\begin{lemma}[Calibration]\label{lem:calibration}
    Suppose a distributed state $\Phi_{S_1...S_n}$ can be used to complete an ESS with unknown partner, in the weak sense that the protocol returns a maximally entangled state $\Psi^i$ for each authorized pair $A_i$, but which maximally entangled state may be different for each pair.
    Then, the same distributed state $\Phi_{S_1...S_n}$ can be used with a modified protocol to always return the same state $\ket{\Psi^+}$ for every pairing $A_i$. 
\end{lemma}
\begin{proof}\,
    For each connected component of $G_\A$, consider a spanning tree and pick a root, then label the vertices of that component by their depth. 
    We can unitarily conjugate the distillation channels according to the tree structure so that all edges in the tree are calibrated in the sense that they indeed give rise to the state $\ket{\Psi^+}$.
    In particular, we leave the recovery channel corresponding to the root unchanged. 
    Then at the first level, if an edge $(j,k)$ ($j$ the root) corresponds to a returned maximally entangled state $\ket{\Psi_{j,k}}_{X_jX_k}=U^{j,k}_{X_k}\ket{\Psi^+}_{X_jX_k}$, we modify the recovery channel acting on $T_k$ from $\mathcal{N}_{T_k\rightarrow X_k}$ to $(U_{X_k}^{j,k})^\dagger\mathcal{N}_{T_k\rightarrow X_k}(\cdot)U_{X_k}^{j,k}$, so that the returned state is now $\ket{\Psi^+}_{X_jX_k}$ as needed. 
    In the next layer, if the returned state is initially $U^{k,\ell}_{X_\ell}\ket{\Psi^+}_{X_kX_\ell}$ we conjugate the recovery channel by $(U^{j,k}_{X_\ell})^T(U^{k,\ell}_{X_\ell})^\dagger$ to both undo the correction from the previous layer and make the needed correction for the current layer. 
    Proceeding in this way we can correctly return $\ket{\Psi^+}$ for every authorized pair corresponding to an edge in the spanning tree. 

    Since not all edges in the given connected component of $G_\A$ are in the spanning tree, this has only calibrated a subset of the authorized pairs.
    To handle the remaining authorized pairs use remark \ref{remark:generalizednocycles} so that $G_\A$ must contain no odd cycles.
    This also means $G_\A$ is bipartite.
    This means the edges missing from the spanning tree will never connect layers of the tree with the same parity. 
    Considering such an edge, there is an odd length path through the spanning tree with the same start and end points. 
    But then the proof of \cref{lemma:transitiveESS} gives that these pairs are automatically authorized, and in fact authorized in that they return $\ket{\Psi^+}$ as needed. 
\end{proof}

\subsection{Characterization without unauthorized pairs}\label{sec-ESSunknownsimple}

In this section we give a characterization of entanglement sharing schemes with an unknown partner in the case where there are no unauthorized pair constraints: we only require there to be a set of authorized pairs. 
Note that it can still be the case that many pairings are unauthorized, but we do not put this in as a requirement.
For the known partner case, this setting is trivial: we can share an EPR pair between every authorized pair and make any collection of pairs authorized this way. 
For the unknown partner case the setting is still non-trivial however. 

In this setting, the maximality related conditions trivialize, since they specify that certain additional pairs must be authorized given a set of authorized pairs, and this is always allowed when not specifying unauthorized pairs. 
In contrast, the no odd cycles and monogamy conditions remain non-trivial: these are conditions on collections of authorized pairs needed for those pairs to all be made authorized (not statements that additional pairs must be authorized). 
This reasoning leads us to the following theorem. 

\begin{theorem}\label{thm:ESSwithunknownpartner}
A pair access structure $\mathcal{S}=(\mathcal{A}, \varnothing)$ can be realized as an entanglement sharing scheme with an unknown partner if and only if the following conditions hold. 
\begin{itemize}
    \item \textbf{No odd cycles:} The graph $G_{\mathcal{A}}$ must not contain any odd cycles. 
    \item \textbf{Monogamy:} Suppose there is an even length path $\{T_1,T_2\}$, $\{T_2,T_3\}$, ...,$\{T_{k-1},T_k\}$ in $G_\mathcal{A}$. Then $T_1\cap T_k\neq \emptyset$. 
\end{itemize}
\end{theorem}

\begin{proof}\,
    Necessity has been proved as \cref{lemma:nooddcyclesESS} and \cref{lemma:monogamyESSWKP}.
    It remains to prove sufficiency. 
    We will first construct a scheme that addresses a single connected component, and then combine these constructions. 

    Let the graph $G_\mathcal{A}$ 
    have connected components, $G_\mathcal{A}^a$, each indexed by the label $a$.     
    Consider an arbitrary $G_\mathcal{A}^a$.  
    The no odd cycles condition and \cref{lemma:oddcyclesandbipartite} imply that the graph $G_\mathcal{A}^a$ is bipartite. 
    Call the two vertex sets induced by the bipartition $\mathcal{V}_L^a$ and $\mathcal{V}_R^a$. 
    Our strategy will be to take a single copy of $\ket{\Psi^+}_{A_LA_R}$, and share $A_L$ among the vertices in $\mathcal{V}_L^a$ and $A_R$ among the vertices in $\mathcal{V}_R^a$ using appropriate quantum secret sharing schemes (see \cref{sec:QSS}). 
    That is, we want a QSS for $A_L$ so that each vertex in $\mathcal{V}_L^a$ is an authorized set, similarly for $A_R$ and $\mathcal{V}_R^a$.  
    From \cref{sec:QSS}, it suffices to show $\mathcal{V}_L^a$ satisfies no-cloning and monotonicity for QSS (defined right after \cref{eq:lastlinesignshare}).  

    We first show that $\mathcal{V}_L^a$ satisfies the no-cloning property.
    To see this, consider $T_1, T_2\in \mathcal{V}_L^a$. 
    Because we have limited ourselves to a single connected component, there is at least one path between $T_1$ and $T_2$. 
    Since the graph is bipartite and $T_1,T_2\in \mathcal{V}_L^a$, this path must have even length. 
    But then monogamy for ESS (\cref{lemma:monogamyESSWKP})    
    implies $T_1\cap T_2\neq \emptyset$, as needed. 
    To obtain monotonicity for the QSS, we add vertices to $\mathcal{V}_L^a$.  In particular, add all supersets of each $T\in \mathcal{V}_L^a$, and obtain $\bar{\mathcal{V}}_L^a$. Note that $\bar{\mathcal{V}}_L^a$ is the minimal set that satisfies monotonicity and includes $\mathcal{V}_L^a$. 
    It is clear that $\bar{\mathcal{V}}_L^a$ also satisfies no-cloning.
    Now, $\bar{\mathcal{V}}_L^a$ satisfies monotonicity and no-cloning, so forms a valid collection of authorized sets for a quantum secret sharing scheme.
    We then use a QSS to share $A_L$ among $\bar{\mathcal{V}}_L^a$.
    Similarly we use a QSS to share $A_R$ among $\bar{\mathcal{V}}_R^a$. 
    
    We repeat this procedure for each connected component and then include all of the constructed states together to obtain the final construction.
    In this final state $\rho_\A$, each pair $\{T_i, T_j\}\in\mathcal{A}$ will be authorized: explicitly, the maps $\mathcal{N}_{T_i}$ are given as follows: We consider the graph $G_\mathcal{A}$ and find which (unique) connected component $T_i$ belongs to.
    Then $\mathcal{N}_{T_i}$ is defined as the recovery channel for the corresponding QSS.
\end{proof}

\subsection{General access structures for stabilizer states}\label{sec-stabiffunknown}

In this section we provide necessary and sufficient conditions for an access structure to be realized in stabilizer constructions. 
We work with qubit systems for simplicity. 
The generalization to other prime-dimensional qudits is straightforward.

Consider an ESS with unknown partners, defined by a state and family of decoding channels, $(\rho,\{\N_T\})$.
We may realize each distillation channel $\N_T$ as a unitary $U_T$ followed by a partial trace:
$\N_T(\cdot)=\tr_{(o_T)^c}(U_T(\cdot)U_T^\dagger)$, where $o_T$ is the output qubit of channel $\N_T$.
Define operators $X_T=U_T^\dagger (X_{o_T}\otimes{I}) U_T$, where $X_{o_T}$ is the Pauli-X operator on qubit $o_T$, and ${I}$ is the identity operator on $T\setminus o_T$.
Define $Z_T$ similarly.
Since we focus on constructions in the stabilizer formalism, we always assume $U_T$ are Clifford unitaries, which means that also $X_T$ and $Z_T$ are Pauli operators.
We have that $\N_T\otimes \N_{T'}$ outputs the EPR state for $\forall \{T,T'\}\in\A$.
Namely, $X_{o_T}X_{o_{T'}}$ and $Z_{o_T}Z_{o_{T'}}$ are stabilizers of the output EPR state.
Accordingly, $X_TX_{T'}$ and $Z_TZ_{T'}$ are stabilizers of the original state $\rho$.

We will first establish some facts about these stabilizers. 
\begin{lemma}\label{lem:bgstab}
Consider a path of vertices $T_1, T_2, \cdots$ in the graph $G_{\A}$. 
Define $X_{T_i}$ and $Z_{T_i}$ as in the last paragraph.
Then: 
\begin{enumerate}
    \item $\forall i\neq j$, $X_{T_i}X_{T_j}$ and $Z_{T_i}Z_{T_j}$ are stabilizers of $\rho$.
    \item We have the following (anti-)commutation relations:
    \begin{equation}
        \forall i,k,\,\,\,\,\,\,\{X_{T_i},Z_{T_{i+2k}}\}=0, \,\,\,\,[X_{T_i},Z_{T_{i+2k+1}}]=0.
    \end{equation}
\end{enumerate}
\end{lemma}
\begin{proof}\,
(1) Since each $X_{T_i}X_{T_{i+1}}$ is a stabilizer of $\rho$, so is $X_{T_i}X_{T_j}=\prod_{k=i}^{j-1}X_{T_k}X_{T_{k+1}}$.
Similar for $Z$ stabilizers.

(2) We work by induction on $k$.
    For $k=0$, we need to prove $\{X_{T_i},Z_{T_i}\}=0$ and $[X_{T_i},Z_{T_{i+1}}]=0$ for all $i$.
    The first equation follows from the definition of $X_T$ and $Z_T$; the second equation follows from $T_{i}\cap T_{i+1}=\emptyset$. 
    Now let us consider $k\geq 1$.
    We have $[X_{T_i}X_{T_{i+1}},Z_{T_{i+2k-1}}Z_{T_{i+2k}}]=0$, since both are stabilizers due to (1).
    By the induction hypothesis, $[X_{T_i},Z_{T_{i+2k-1}}]=[X_{T_{i+1}},Z_{T_{i+2k}}]=0$, $\{X_{T_{i+1}},Z_{T_{i+2k-1}}\}=0$.
    It follows that $\{X_{T_i},Z_{T_{i+2k}}\}=0$. 
    A similar argument gives $[X_{T_i},Z_{T_{i+2k+1}}]=0$. 
\end{proof}

We note that in the above lemma, (1) readily implies transitivity (\cref{lemma:transitiveESS}).
Indeed, if both $X_TX_{T'}$ and $Z_TZ_{T'}$ are stabilizers, and $T\cap T'=\emptyset$, then $\N_T\otimes\N_{T'}$ recovers an EPR pair.
Moreover, (2) readily implies both monogamy (\cref{lemma:monogamyESSWKP}) and no odd cycles (\cref{lemma:nooddcyclesESS}).
Indeed, suppose there is an even-length path $T_1, T_2, \cdots, T_{2k+1}$, then $\{X_{T_1},Z_{T_{2k+1}}\}=0$ due to (2), implying $T_1\cap T_{2k+1}\neq \emptyset$, which is exactly the claim of monogamy.
The claim of no odd cycle, that $\{T_{2k+1},T_1\}\notin\A$, also follows. Otherwise, applying (2) to $\{T_{2k+1},T_1\}\in\A$, we would get $[X_{T_1},Z_{T_{1+2k}}]=0$, a contradiction.

Now we derive an equivalent condition for stabilizer ESS with unknown partners.
Let us consider a connected component of $G_\A$, which recall is a bipartite graph by the no odd cycles condition.
For every pair of vertices (not necessarily connected) $T_p$ and $T_q$, we consider $V=X_{T_p}$, $W=Z_{T_q}$ and define an $n$-dimensional vector $u_{pq}\in \mathbb{F}^{n}_2$ to record their commutation relation:
\begin{equation}
    ({-1})^{(u_{pq})_s}=V_{s}W_{s}V_{s}^{-1}W_{s}^{-1}.
\end{equation}
Here $s$ is the index of the vector $u_{pq}$ corresponding to the label of parties, $V_s$ is the tensor factor (a Pauli matrix) of $V$ on the $s$-th party, and $W_s$ is the tensor factor of $W$ on the $s$-th party.
It is immediate that,
\begin{equation}\label{eq:KPrank1}
    \supp(u_{pq})\subseteq T_p\cap T_q.
\end{equation}
From the (anti-)commutation relation (\cref{lem:bgstab}), we have
\begin{equation}\label{eq:KPrank2}
   z_0\cdot u_{pq}=
   \begin{cases}
       1&~\text{if $T_p$ and $T_q$ belong to the same side of~} G_\A\\
       0&~\text{otherwise}
   \end{cases}.
\end{equation}
Here $z_0\in\mathbb{F}^{n}_2$ is the vector with all entries equal to 1.

Furthermore, given any four subsets $T_p,T_q,T_r,T_s$, we could consider $X_{T_p}X_{T_r}$ and $Z_{T_q}Z_{T_s}$, which are stabilizers due to \cref{lem:bgstab}.
Their commutation relation is encoded in $u_{pq}+u_{ps}+u_{rq}+u_{rs}$.
For any $R\notin\cup\A$, consider $\{R,R^c\}$.
Due to \cref{lemma:indistillable}, which says that no EPR pair can be distilled, and \cref{remark:tripartitestabilizer}, which then implies the state must be separable across a bipartition with no EPR pairs, the state $\rho$ must be separable with respect to the pair $\{R,R^c\}$.
We can now re-run the argument in the proof of \cref{lemma:rank}, around \cref{eq:rank1.1}: let $x=u_{pq}+u_{ps}+u_{rq}+u_{rs}$ be the vector describing the commutation properties of the two stabilizers. 
Then we found that for any unauthorized pair $U$, we have $z_Ux=0$. 
In our setting we see that $\{R,R^c\}$ is an unauthorized pair, so that
\begin{equation}\label{eq:KPrank3}
    \forall R\notin\cup\A, \quad z_{R}\cdot (u_{pq}+u_{ps}+u_{rq}+u_{rs})=0.
\end{equation}
Here $z_{R}\in\mathbb{F}^{n}_2$ is the vector with all components in $R$ equal to 1 (and 0 on $R^c$).

The three relations \cref{eq:KPrank1}, \eqref{eq:KPrank2}, and \eqref{eq:KPrank3} can be viewed as equations for variables $\{u_{pq}\}$. 
We have shown that the existence of a solution for these equations\footnote{If we ignore \cref{eq:KPrank3}, which reflects the maximality requirement, \cref{eq:KPrank1,eq:KPrank2} admit a solution if and only if the monogamy condition holds for both sides of $G_{\mathcal{A}}$. Similar consideration gives rise to an alternative construction of the ordinary quantum secret sharing, see \cref{app-QSS}.}, which can be expressed as a rank condition similar to \cref{lemma:rank}, is necessary for $\A$ to be a valid access structure. 
We further claim that these conditions, along with no odd cycles and transitivity, are sufficient. 

\begin{theorem}\label{thm:ESSunknown}
    ESS with an unknown partner, and using only stabilizer encodings and decodings,\footnote{Note that we do not assume an adversary is using only Clifford decodings; the statement is that on authorized pairs, the players can recover using only Clifford operation.} can realize access structure $(\mathcal{A},\mathcal{U})$, $\A$ maximal, with strong security if and only if the following conditions hold. 
    \begin{itemize}
        \item \textbf{No odd cycles:} $G_{\mathcal{A}}$ must not contain an odd length cycle.
        \item \textbf{Transitivity:} If there is a path of edges $\{T_1,T_2\}$,...,$\{T_{k-1},T_{k}\}\in G_{\mathcal{A}}$ with odd length and $T_1\cap T_{k}=\emptyset$, then $\{T_1,T_{k}\}$ is authorized.
        \item There exists a solution of \cref{eq:KPrank1}, \eqref{eq:KPrank2}, and \eqref{eq:KPrank3} for each connected component of $G_\A$.
    \end{itemize}
\end{theorem}

Before proceeding to prove this, notice that the monogamy condition and the weak monotonicity condition, which we have shown to be necessary, are missing here.
In fact, they are implied by the conditions stated in the above theorem.

First, no odd cycles and \cref{eq:KPrank1,eq:KPrank2} imply monogamy.
Suppose there is an even-length path from $T_p$ to $T_q$, then they must belong to the same side in the bipartite graph $G_\A$.
In particular, \cref{eq:KPrank2} implies $u_{pq}\neq 0$, which further implies $T_p\cap T_q\neq \emptyset$ due to \cref{eq:KPrank1}.

Second, the above conditions also imply weak monotonicity (\cref{lemma:weakmono}).
In fact, consider the setting for weak monotonicity: $\{T_1, T_2\}\in \mathcal{A}$, $T_1\subseteq T_3$ and $T_3\cap T_2=\emptyset$.
Applying \cref{eq:KPrank3} to $p=q=1$ and $r=s=2$ and assuming $T_3\notin \cup \A$, we would get
\begin{equation}\label{eq-local-1}
    z_{T_3}\cdot (u_{11}+u_{12}+u_{21}+u_{22})=0.
\end{equation}
On the other hand, since $\supp(u_{11})\subseteq T_1\subseteq T_3$ and using \cref{eq:KPrank2}, we have
\begin{equation}
    z_{T_3}u_{11}=z_0 u_{11}=1,
\end{equation}
where the last equality is due to \cref{eq:KPrank2}.
Further, recall that $\supp(u_{pq})\subseteq T_p\cap T_q$, which means $\supp (u_{12})\subseteq T_2$, and then since $T_2\cap T_3=\emptyset$, $u_{12}$ has no common support with $T_3$, 
\begin{equation}
    z_{T_3}u_{12}=0,
\end{equation}
By the same argument, 
\begin{align}
    z_{T_3}u_{21}=0,\nonumber \\
    z_{T_3}u_{22}=0.
\end{align}
Using the last four equations we see that \cref{eq-local-1} does not hold, which is a contradiction, so we must have $T_3\in \cup \A$, which is exactly the claim of weak monotonicity.

\vspace{0.2cm}
\begin{proof}[Proof of \cref{thm:ESSunknown}]
We have already shown necessity, as \cref{lemma:nooddcyclesESS}, \cref{lemma:transitiveESS}, and the discussion just preceding the theorem statement.
Conversely, assuming these conditions hold, let us construct a stabilizer state $\rho$ that satisfies the ESS property.
It suffices to consider a single connected component of $G_\A$.
The overall state will be the tensor product of the states constructed for each connected component.

By assumption, \cref{eq:KPrank1,eq:KPrank2,eq:KPrank3} admit a solution, denoted as $\{u_{pq}\}$, where $p$ and $q$ are vertices of $G_\A$.
To construct a stabilizer state, we will specify a rule to assign qubits to parties (each party may contain multiple qubits, and the number of qubits each party has may differ) and a rule to define stabilizers, denoted as $\{X_p\}$ and $\{Z_p\}$.
More precisely, they are defined by the following procedure
\begin{itemize}
    \item Initially, there are no qubits; each $X_p$ and $Z_p$ is the identity operator;
    \item For any $p$ and $q$ and any party $s$, if $(u_{pq})_s=1$, we add a qubit to party $s$, then extend $X_p$ by (tensor multiplying it with) the Pauli $X$ of the newly added qubit, and extend $Z_q$ by the Pauli $Z$ of the newly added qubit.
\end{itemize}
After this procedure, each party will contain some (or no) qubits, and $\{X_p\}$ and $\{Z_p\}$ will be Pauli strings. 
Furthermore, the above constructed $\{X_p\}$ and $\{Z_p\}$ satisfy the following properties:
\begin{itemize}
    \item $\supp(X_p)\subseteq T_p$, $\supp(Z_p)\subseteq T_p$ (this is due to \cref{eq:KPrank1});
    \item the commutation relation between $X_p$ and $Z_q$ is given by $u_{pq}$; 
    \item $\forall p,q,r,s, \,\,\,[X_{T_p}X_{T_r},Z_{T_q}Z_{T_s}]=0$; this is due to the previous property and \cref{eq:KPrank2}.
\end{itemize}
Therefore, we may define a (in general, mixed) state by using $X_{T_p}X_{T_r}$ and $Z_{T_q}Z_{T_s}$ as stabilizers.

Taking $p=q$ in the first line of \cref{eq:KPrank2}, we see that $X_p$ and $Z_p$ always anticommute.
Therefore, there exists a Clifford unitary $U_p$ that transforms $X_p$ and $Z_p$ to Pauli $X$ and $Z$.

We claim that the above constructed $\rho$ together with the Clifford unitaries $\{U_p\}$ satisfies the ESS property.
In fact, if $T_p$ and $T_q$ share an edge in $G_\A$, then the stabilizers $X_pX_q$ and $Z_pZ_q$ and the anticommutation relations on $T_p$ and $T_q$ already imply that $U_p$ and $U_q$ distill the EPR state out of $\rho$ (by \cref{lemma:stabdistillcondition}).

It still remains to show maximality.
Consider a pair $\{T_1,T_2\}\notin\A$. There are two possibilities.

(1) $T_1\notin\cup\A$ or $T_2\notin\cup\A$.
Without loss of generality, we assume $T_1\notin\cup\A$.
We claim that $\rho$ is separable with respect to $\{T_1,T_2\}$.
It suffices to prove the separability of $\{T_1,T_1^c\}$.
Notice that, for our state $\rho$, all stabilizers are of the form of (products of) $X_{T_p}X_{T_r}$ and/or $Z_{T_q}Z_{T_s}$, and their commutation relations are linear combinations of $u_{pq}+u_{ps}+u_{rq}+u_{rs}$.
\Cref{eq:KPrank3} then implies there do not exist stabilizers $V$ and $W$ with the commutation relations that correspond to a distillable EPR pair (as characterized in \cref{lemma:stabdistillcondition}).

(2) $T_1, T_2\in\cup\A$. 
In this case, we claim that $T_1$ and $T_2$ must be in different connected components.
Suppose instead they are in the same connected component.
Due to monogamy, they cannot be on the same side (recall that this is a bipartite graph) of a single component (note that, by definition, $T_1\cap T_2=\emptyset$). 
Then due to transitivity, we have $\{T_1,T_2\}\in\A$, which contradicts the assumption that $\{T_1,T_2\}\notin\A$.
It follows that $\{T_1,T_2\}$ cannot be added to $\A$:
we must use existing distillation unitaries $U_{T_1}$ and $U_{T_2}$, but they correspond to different connected components and hence the output qubits are unentangled.
\end{proof}

Before closing this subsection, we make two comments on the conditions \cref{eq:KPrank2,eq:KPrank3}.
First, with \cref{eq:KPrank1} as prerequisite, \cref{eq:KPrank2} is equivalent to the following two equations:
\begin{align}
    \forall& p, \,\,\,\,\qquad \,\,\,\,\, z_0 u_{pp}=1\\
\forall& p,q,r,s, \,\,\,\,\,z_0(u_{pq}+u_{ps}+u_{rq}+u_{rs})=0.
\end{align}
In fact, the equivalence is almost a restatement of \cref{lem:bgstab}.
To prove it, we simply repeat the proof of \cref{lem:bgstab}, while replacing the (anti-)commutation relations with equations for $z_0u_{pq}$.
Second, \cref{eq:KPrank3} is equivalent to that for a spanning tree (of each connected component of $G_\A$) only.
Namely, it suffices to have \cref{eq:KPrank3} for all $(p,r)$ and $(q,s)$ corresponding to edges in a spanning tree.
Other equations then follow from linear combinations.

\section{Non-stabilizer constructions}
\label{sec:nonstab}

In \cref{sec-stabiffknown,sec-stabiffunknown}, we found the equivalent conditions for ESS with known and unknown partners respectively. 
In this section, we discuss realizations via non-stabilizer states.
As a general comment, we note that the extra conditions---the second condition in \cref{thm-stabiffknown} and the third condition in \cref{thm:ESSunknown}---are rooted in the tensor product structure of the Pauli operators, and are specific to the stabilizer formalism.
Therefore, we should not expect them to hold in general.

\subsection{Known partner}\label{sec:knownpartnersketch}

At the beginning of \cref{sec-stabiffknown}, we discussed a minimal obstruction for a monotonic set of authorized pairs to be realized via stabilizer states.
Namely, it is impossible for a tripartite stabilizer state to contain only one authorized pair $\{\{S_1\},\{S_2,S_3\}\}$.
However, we can achieve this with the following non-stabilizer state
\begin{align}
    \ket{\Psi}= \frac{1}{\sqrt{2}}\ket{000} + \frac{1}{2}\ket{110}+\frac{1}{2}\ket{101}.
\end{align}
One can check that the partition $S_1:S_2S_3$ has Schmidt coefficients $(1/\sqrt{2}, 1/\sqrt{2})$ so an EPR pair can be recovered by local unitaries, but the partitions $S_2:S_1S_3$ and $S_3:S_1S_2$ have Schmidt coefficients $(\sqrt{3}/2, 1/2)$ and hence cannot recover an EPR pair.

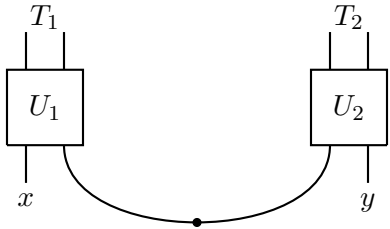
\begin{figure}[h!]
    \centering
    \begin{tikzpicture}[scale=0.5]
    
    \draw[thick] (-5,-5) -- (-5,-3) -- (-3,-3) -- (-3,-5) -- (-5,-5);
    \node at (-4,-4) {$U_1$};
    
    \draw[thick] (5,-5) -- (5,-3) -- (3,-3) -- (3,-5) -- (5,-5);
    \node at (4,-4) {$U_2$};

    \draw[thick] (-4.5,-3) -- (-4.5,-2);
    \draw[thick] (4.5,-3) -- (4.5,-2);

    \draw[thick] (-3.5,-3) -- (-3.5,-2);
    \draw[thick] (3.5,-3) -- (3.5,-2);
    
    \draw[thick] (-3.5,-5) to [out=-90,in=-90] (3.5,-5);
    \draw[black] plot [mark=*, mark size=3] coordinates{(0,-7.05)};

    \node[above] at (-4,-2.2) {$T_1$};
    \node[above] at (4,-2.2) {$T_2$};
    
    \draw[thick] (-4.5,-6) -- (-4.5,-5);
    \node[below] at (-4.5,-6) {$x$};
    
    \draw[thick] (4.5,-6) -- (4.5,-5);
    \node[below] at (4.5,-6) {$y$};
    
    \end{tikzpicture}
    \caption{A construction for weakly secure ESS. Here $U_1$ and $U_2$ are Haar random.}
    \label{fig:haar}
\end{figure}

We will generalize this construction to realize weak security for any access structure satisfying monotonicity. 
Because we are mainly interested in constructions with strong security, we only sketch the construction informally. 
Returning to the above, note that this state can be constructed from an EPR state followed by a non-Clifford gate:
\begin{equation}
    \ket{\Psi}=\frac{1}{\sqrt{2}}(I_1\otimes U_{23})(\ket{00}+\ket{11})\ket{0}=\frac{1}{\sqrt{2}}\left(
    \ket{0}\ket{00}+\ket{1}\frac{\ket{01}+\ket{10}}{\sqrt{2}}
    \right).
\end{equation}
In general, given an authorized pair $\{T_1,T_2\}$, one may start with an EPR state, and then apply a Haar random unitary on each side (see \cref{fig:haar}).
As long as $U_1$ and $U_2$ are generic, only $\{T_1,T_2\}$ is exactly authorized. 
That is, all other pairs are unauthorized in the sense that one cannot recover an EPR state exactly.
For any access structure $\A$, one may repeat the above procedure and simply stack (tensor product) the constructions.
While stacking multiple states may lead to better EPR distillation for pairs not in $\A$, at least such recovery cannot be exact.

While we initialize the ancilla qubit as a pure state $\ket{0}$ in the above example, we could also initialize it as a maximally mixed state.
In fact, we expect this makes the construction more secure.
Let us consider the case $|T_2|=1$ as an example so we do not need $U_2$ and $y$.
If we set $x$ as a pure state, then the overall state of $T_1\cup T_2$ can be understood as a Haar random state. 
Due to Page's theorem \cite{Page:1993df}, a small subset of $T_1$ will be nearly maximally entangled with the complement, although not perfectly.
On the other hand, if we set $x$ to be maximally entangled, then it follows that $x\cup T_1\cup T_2$ can be understood as a Haar random state via a Choi–Jamiołkowski isomorphism. 
Due to \cite{li2025tripartite}, after choosing the dimensions carefully, there will be nearly no distillable (using only local operations) EPR states between $A$, a small subset of $T_1$, and $(T_1\cup T_2)\backslash A$.

To summarize, for any access structure satisfying monotonicity only, we have constructed an ESS where authorized pairs can recover exact EPR states while unauthorized pairs cannot. 
However, unauthorized pairs may be able to recover states $\epsilon$-close to EPR pairs, where $\epsilon$ becomes small as the number of authorized pairs becomes large. 
Thus this construction satisfies weak security (\cref{def:weaksecurity}) but not strong security (\cref{def:strongsecurity}).
We have not achieved a construction with strong security. 
However, we make the following conjecture. 

\begin{conjecture}
    ESS with a known partner, strong security, and access structure $\mathcal{S}$ can be realized if and only if $\mathcal{S}$ satisfies monotonicity. 
\end{conjecture}

We leave further consideration of this conjecture to future work. 

\subsection{Unknown partner}\label{sec:unknownsketch}

In \cref{sec:neces-ESS-unknown}, we proved several necessary conditions for ESS with unknown partners:
no odd cycles, monogamy, transitivity, and weak monotonicity.
In this subsection, we argue that these conditions are also sufficient to guarantee the existence of ESS with unknown partners and weak security.
The construction shares several features with the above subsection: both are generally non-stabilizer states, both employ non-Clifford unitaries to break the tensor product structure, and both achieve weak but not necessarily strong security. 
Again because we are ultimately interested in constructions with strong security, we leave the construction informal.

The construction is based on \cref{sec-ESSunknownsimple} and \cref{app-QSS} (we recommend reading them before reading this subsection).
Following \cref{sec-ESSunknownsimple}, we work on each individual connected component of $G_\A$.
The bipartition construction and the monogamy condition induce two valid QSS structures, $\bar{\mathcal{V}}_L$ and $\bar{\mathcal{V}}_R$, one for each side.
We then encode each side of an EPR state via a QSS.
The resulting state is an ESS with unknown partner such that all pairs in $\A$ are authorized exactly.
Denote the corresponding distillation channels as $\{\N_T\}$.

To rule out other pairs that may be authorized accidentally, we need to specify the QSS in detail.
The construction will be a further modification of \cref{app-QSS}.
We will first follow the procedure there.
Notice that there are three types of qubits: those assigned in the first step  (initialization), second step (anticommutation-1), and third step (anticommutation-2).
We then apply generic (Haar random) unitaries on those qubits of the first step.
More precisely, for each $T\in (\bar{\mathcal{V}}_L)_{\min}$ (and similarly for the right side), we apply a generic unitary on the $|T|$ qubits that were assigned during the initialization procedure corresponding to $T$. 

The non-Clifford unitaries break the tensor product structure of the stabilizers.
As a result, $\{T_1,T_2\}$ can distill an exact EPR state only if there exists $\{T_3, T_4\}\in\mathcal{A}$ such that (i) $T_3\subseteq  T_1$, $T_4\subseteq T_2$, and (ii) $\mathcal{N}_{T_1}\otimes \mathcal{N}_{T_2}$ distills the EPR pair that was encoded through two-sided QSS corresponding to the connected component containing both $T_3$ and $T_4$.

Now let us prove that every pairing $\{T_1,T_2\}$ which is authorized by this construction is actually already a pairing in $\A$. 
Suppose the contrary: that $\{T_1, T_2\}$ is authorized with $T_1\cap T_2 =\emptyset$ and $\{T_1, T_2\}\notin\mathcal{A}$.
We will use that $\A$ is a valid set of authorized pairs, meaning it satisfies the necessary conditions of \cref{sec:neces-ESS-unknown}: no odd cycles, monogamy, transitivity, and weak monotonicity.
It follows from the above paragraph that there exists $\{T_3, T_4\}\in\mathcal{A}$ satisfying the two conditions stated there.
By (i), we get that $T_1\cap T_4 = T_2\cap T_3=\emptyset$.
This makes $\{T_1,T_4\}$ and $\{T_2,T_3\}$ legitimate pairings, and if they were both in $\A$ then using that $\{T_1, T_4\}$, $\{T_4, T_3\}$, $\{T_3, T_2\}$ are all authorized, transitivity of $\A$ would imply $\{T_1,T_2\} \in \A$, a contradiction. 
Thus one of $\{T_1, T_4\}$ and $\{T_3, T_2\}$ must not be in $\A$; without loss of generality, we assume $\{T_1,T_4\}\notin\mathcal{A}$.
We claim this implies that $T_1$ and $T_4$ do not belong to the same connected component of the graph $G_\mathcal{A}$. 
Assume they are in the same connected component, meaning there is a path between them.
Then if $T_1$ and $T_4$ are connected by an even-length path, monogamy would imply that $T_1\cap T_4\neq\emptyset$, contradicting $T_1\cap T_4=\emptyset$.
Meanwhile if they are connected by an odd-length path, transitivity and $T_1\cap T_4=\emptyset$ would imply that $\{T_1, T_4\}\in\A$, contradicting $\{T_1, T_4\}\notin\A$.
Thus we can conclude $T_1,T_4$ are in distinct connected components. 

On the other hand, since $T_3\subseteq T_1$, $\{T_3, T_4\}\in\mathcal{A}$, $T_1\cap T_4 =\emptyset$, weak monotonicity implies $T_1\in\cup\A$. 
(Namely, there exists $T_5$ such that $\{T_1,T_5\}\in\mathcal{A}$.)
Therefore, $T_1$ and $T_5$ are connected in $G_\mathcal{A}$, and belong to a different connected component from that of $T_3$ and $T_4$. 
Consequently, $\mathcal{N}_{T_1}$ cannot distill one side of the EPR pair that $\mathcal{N}_{T_3}\otimes\mathcal{N}_{T_4}$ distills out. This contradicts our choice of $T_3$ and $T_4$, completing the argument.

We also give the following conjecture, which we leave to future work. 
\begin{conjecture}
    ESS with unknown partner, strong security, and access structure $\mathcal{S}$ can be realized if and only if $\mathcal{S}$ satisfies no odd cycles, monogamy, transitivity, and weak monotonicity. 
\end{conjecture}

\section{Lower bounds on share size}\label{sec:lowerbounds}

We briefly give some lower bounds on share size in entanglement sharing schemes.
The first lower bound only applies in the known partner case (since the setup is forbidden in the unknown partner case), but the second lower bound strategy applies to both. 
Note that we do not expect these lower bounds to fully characterize the achievable efficiency in ESS, since even in QSS (which is a special case) this is not known. 

\vspace{0.2cm}
\noindent \textbf{Multiple pairings:} Suppose that a set $T_1$ is one end of an authorized pair. 
Then it is clear $d_{T_1}\geq d_E$ for $d_E$ the dimension of the shared entangled state $\Psi^+$. 
More is also true: if $T_1$ is one end of $t$ authorized pairs, it must have dimension $d_{T_1}\geq d^t_E$. 
We prove this in the next lemma. 

\begin{lemma}[Degree lower bound]\label{lem:degree-lb}
Consider an ESS storing EPR pairs of dimension $d_E$.
Let a subset $T$ be an endpoint of $t$ authorized pairs, i.e. 
$\{T,T_1\}$, $\dots$, $\{T,T_t\}\in \mathcal{A}$, where the partners $T_1,\dots,T_t$ are pairwise disjoint and disjoint from $T$.
Then
\[
  S(T)\;\ge\; t\,\log d_E.
\]
\end{lemma}

\begin{proof}\,
Fix $j\in\{1,\ldots,t\}$. 
Since $\{T,T_j\}$ is authorized, there is a pair of local operations recovering a perfect $d_E$-dimensional maximally entangled state.
Thus we have that
\begin{align}
    \log d_E \leq E_{\text{sq}}(T:T_j) .
\end{align}
Monogamy of $E_{\mathrm{sq}}$ gives
\begin{equation}\label{eq:Monogamy}
  \sum_{j=1}^{t} E_{\mathrm{sq}}(T{:}T_j)\;\le\; E_{\mathrm{sq}}\!\bigl(T{:}T_1\cdots T_t\bigr).
\end{equation}
It is also true however that
\begin{equation}\label{eq:EsqUpper}
  E_{\mathrm{sq}}(A{:}B)\;\le\;\tfrac12\, I(A{:}B)\;\le\; S(A).
\end{equation}
Combining the last three statements, we have
\[
  t\,\log d_E
  \;\le\; \sum_{j=1}^{t} E_{\mathrm{sq}}(T{:}T_j)
  \;\le\; E_{\mathrm{sq}}\!\bigl(T{:}T_1\cdots T_t\bigr)
  \;\le\; S(T),
\]
as claimed.
\end{proof}

Note that the bound can be saturated: suppose $\{T,T_1\},...,\{T,T_t\}$ are all authorized and should be able to recover $d_E$-dimensional maximally entangled qudit pairs, in which case the bound says $S(T)\geq t\log d_E$. 
Then we can achieve this by distributing $t$ independent $d_E$-dimensional EPR pairs, one for each authorized pair, in which case then $S(T)=t\log d_E$ as needed.

\vspace{0.2cm}
\noindent \textbf{Lower bound on significant shares:} Another lower bound is imposed by a share being ``significant'', in a sense we adapt from the secret sharing literature \cite{gottesman2000theory, imai2003quantum}. 

\begin{definition}
    A share $S_i$ is significant for a pair access structure $\mathcal{S}$ if there exists an unauthorized pair $U=\{T_1,T_2\}$ such that $A=\{S_i\cup T_1,T_2\}$ is authorized.
\end{definition}

We can show a lower bound on the size of significant shares in schemes that satisfy a stronger condition on unauthorized pairs, which is that their reduced density matrices be separable. 
Recall that this is the case in the explicit constructions we have given, so this lower bound applies in those cases. 
We claim that in this setting any significant share $S_i$ must have $\log d_{S_i}\geq \frac{1}{2}\log d_E$, where $d_E$ is the (one-sided) dimension of maximally entangled state distilled from the pair $\{S_i\cup T_1,T_2\}$.

To see this, first note that since we can distill a $d_E$-dimensional maximally entangled state, we have
\begin{align}
    \log d_E & \leq E_{R}(S_iT_1:T_2). 
\end{align}
Then from \cref{eq:SRcontinuity} we have
\begin{align}
    E_{R}(S_iT_1:T_2) \leq E_{R}(T_1:T_2) + 2\log d_{S_i}
\end{align}
Since by assumption the state on $T_1T_2$ is separable, $E_{R}(T_1:T_2)=0$, so we obtain $\log d_E\leq 2\log d_{S_i}$ as needed.

Note that it is not clear if this bound is tight: it is easy to construct cases where $\log d_E=\log d_{S_i}$, but we have not found examples saturating $\frac{1}{2}\log d_E=\log d_{S_i}$. 

We note that the lower bound on significant shares does not hold under the weaker requirement for unauthorized pairs, namely, demanding the inability to distill EPR states only. 
A counterexample is provided by a Haar random state.
Consider an $n$ qubit Haar random state and divide the qubits into three subsystems $T_1, T_2, S$ such that $|T_1|=|T_2|=\frac{1-\epsilon}{2}n$ and $|S|=\epsilon n$, where $\epsilon$ is an arbitrarily small positive constant.
Under this partition, with high probability, almost $\frac{n}{2}$ EPR pairs can be distilled from $S\cup T_1$ and $T_2$, making $\log d_E\approx \frac{n}{2}$, but no EPR states can be distilled (using local operation) from $T_1$ and $T_2$ \cite{li2025tripartite}.

\section{Application to entanglement summoning}\label{sec:summoning}

In this section we describe an application of the theory of entanglement sharing to understanding an open problem in entanglement summoning \cite{adlam2018relativistic,dolev2021distributing}. 
Without introducing the general context of entanglement summoning, we pose a particular problem from that context.
This problem was the simplest uncharacterized example appearing in entanglement summoning, so resolving it is of general interest for that theory. 
As well, the problem can be framed in practical terms and may be of independent interest. 

\begin{figure}
\centering
\begin{subfigure}{0.49\textwidth}
\centering
\begin{tikzpicture}[scale=1.6,rotate=19]
\begin{scope}[thick,decoration={markings, mark=at position 0.5 with {\arrow{triangle 45}}}] 
  
\draw[fill=black] (1,0) circle (0.05cm);
\node[right] at (1,0) {$D_1$};

\draw[fill=black] (0.309,0.951) circle (0.05cm);
\node[above] at (0.309,0.951) {$D_2$};

\draw[fill=black] (-0.809,0.588) circle (0.05cm);
\node[left] at (-0.809,0.588) {$D_3$};

\draw[fill=black] (-0.809,-0.588) circle (0.05cm);
\node[below left] at (-0.809,-0.588) {$D_4$};

\draw[fill=black] (0.309,-0.951) circle (0.05cm);
\node[below right] at (0.309,-0.951) {$D_5$};

\begin{scope} [rotate=0]
\draw (1,0) -- (0.309,0.951);
\draw[-triangle 45] (1,0) -> (0.481,0.713);
\draw[-triangle 45] (0.309,0.951) -> (0.827,0.238);
\end{scope}

\begin{scope} [rotate=72]
\draw (1,0) -- (0.309,0.951);
\draw[-triangle 45] (1,0) -> (0.481,0.713);
\draw[-triangle 45] (0.309,0.951) -> (0.827,0.238);
\end{scope}

\begin{scope} [rotate=72*2]
\draw (1,0) -- (0.309,0.951);
\draw[-triangle 45] (1,0) -> (0.481,0.713);
\draw[-triangle 45] (0.309,0.951) -> (0.827,0.238);
\end{scope}

\begin{scope} [rotate=72*3]
\draw (1,0) -- (0.309,0.951);
\draw[-triangle 45] (1,0) -> (0.481,0.713);
\draw[-triangle 45] (0.309,0.951) -> (0.827,0.238);
\end{scope}

\begin{scope} [rotate=72*4]
\draw (1,0) -- (0.309,0.951);
\draw[-triangle 45] (1,0) -> (0.481,0.713);
\draw[-triangle 45] (0.309,0.951) -> (0.827,0.238);
\end{scope}

\end{scope}
\end{tikzpicture}
\caption{}\label{fig:pentagonmain}
\end{subfigure}
\begin{subfigure}{0.49\textwidth}
\centering
\begin{tikzpicture}[scale=0.85]

\draw[blue,thick] (0,0) -- (0,3);
\draw[blue,thick] (3,0) -- (3,3);
\draw[blue,thick] (6,0) -- (6,3);

\draw[fill=black] (0,0) circle (0.05cm); 
\node[below] at (0,0) {$c_1$};

\draw[fill=black] (3,0) circle (0.05cm);
\node[below] at (3,0) {$c_2$};

\draw[thick, mid arrow] (0,0) -- (3,3);
\draw[thick, mid arrow] (3,0) -- (0,3);

\draw[fill=black] (3,3) circle (0.05cm); 
\node[above] at (3,3) {$r_2$};

\draw[fill=black] (0,3) circle (0.05cm);
\node[above] at (0,3) {$r_1$};

\draw[fill=black] (6,0) circle (0.05cm);
\node[below] at (6,0) {$c_3$};

\draw[fill=black] (6,3) circle (0.05cm);
\node[above] at (6,3) {$r_3$};

\draw[thick, mid arrow] (3,0) -- (6,3);
\draw[thick, mid arrow] (6,0) -- (3,3);

\draw[dashed] (6,0) -- (7,1);
\draw[dashed] (6,3) -- (7,2);

\draw[dashed] (0,0) -- (-1,1);
\draw[dashed] (0,3) -- (-1,2);

\end{tikzpicture}
\caption{}\label{fig:labscommunication} 
\end{subfigure}
\caption{a) A graph representing a network of 5 labs, connected in a ring configuration by quantum communication channels. b) Each lab is here represented by one of the vertical blue lines. We limit neighbouring labs to a single, simultaneous round of communication. At some initial time, corresponding to the points $c_i$, an input is given at exactly two labs requesting the maximally entangled state be prepared across those two labs. The communication shown, plus an arbitrary pre-shared entangled state, can be used to try to satisfy the request. The two subsystems of the entangled state should be held at the two $r_i$ corresponding to the requests.}
\end{figure}
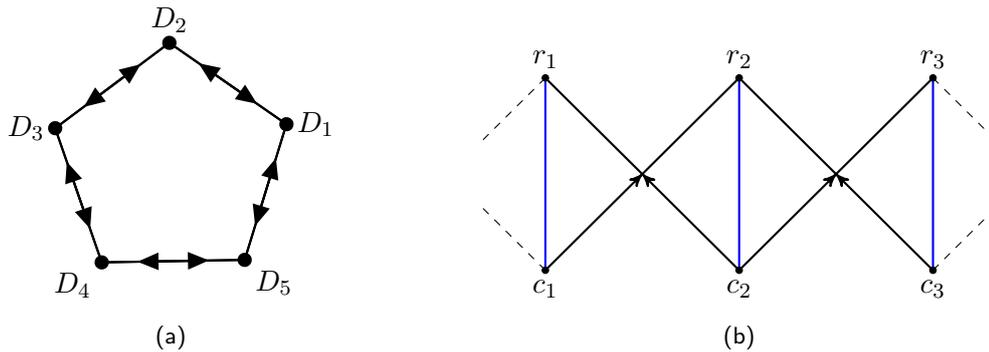

The setting we consider is shown in \cref{fig:pentagonmain}. 
The 5 vertices labelled $D_i$ represent labs; the bidirectional edges between them represent quantum communication links between the labs. 
We allow only a single, simultaneous, round of communication between neighbouring labs. 
This is illustrated in \cref{fig:labscommunication}. 
Importantly, the communication is constrained such that, for example, it is not possible for $D_i$ to signal $D_{i+1}$ who then returns a message to $D_i$, or forwards a message further to $D_{i+2}$.
Such a constraint appears naturally in some applications of quantum networks which rely on timing constraints (e.g. in quantum position-verification \cite{kent2011quantum}), or might be imposed on us by technological constraints. 
For instance, short memory times in quantum memories may mean outputs must be produced before there is time for signals to propagate around the ring of quantum communication links.

Using the network in \cref{fig:pentagonmain}, we want to complete the following task. 
At exactly two of the labs, requests are given simultaneously to produce a maximally entangled pair across the two labs. 
It is necessary to specify a single subsystem which is maximally entangled with the system specified at the other requested lab\footnote{In other words one cannot simply return a large system, composing many parts, each of which is entangled with one of the other labs.}. 
We formalize this by saying bits $b_i$ are given at a spacetime location $c_i$, and exactly two of the $b_i$ will be $1$, indicating entanglement should be prepared across those two labs.
We can then use an arbitrary pre-shared entangled state, plus the limited communication afforded by the network, to try to satisfy this request.
The entangled quantum systems should be returned (to a referee) at a later time near spacetime point $r_i$ (see \cref{fig:labscommunication}).
The points $c_i$ and $r_i$ together define the location of the lab $D_i$. 
In the language of \cite{adlam2018relativistic, dolev2021distributing}, we are asking if a maximally entangled state can be ``summoned'' to the network defined by the pentagon. 

Note that this task cannot be completed by simply sharing EPR pairs between each pair of labs: if one did so, given a request at $D_i$ we have only partial information about the requests at the remaining labs, so although we hold a system entangled with every other lab, it is not clear which system to return. 
In fact, we claim that the summoning problem shown in \cref{fig:pentagonmain} is impossible to implement perfectly by any protocol. 
In fact, we will show that completing it implies implementing an entanglement sharing scheme with unknown partner that has an odd cycle in its authorized pair graph, which recall we showed is impossible in \cref{lemma:nooddcyclesESS}. 

We consider the pentagon task in the restricted setting where only non-neighbouring labs get calls; we will show this is impossible to do perfectly, and hence the harder case (of unrestricted calls) is also impossible. 
A general strategy for completing this task involves beginning with a pre-shared quantum state $\ket{\Psi}_{X_1X_2X_3X_4X_5}$, with system $X_i$ held at $D_i$. 
Then, upon receiving the bit $b_i$, at each lab we act locally on $X_i$ with a quantum operation, then send output systems to $D_{i-1}$ and $D_{i+1}$, and keep another output subsystem at $D_i$.
These operations can depend on the value of the bit $b_i$ given at $c_i$. 
We can notice that without loss of generality, when there is a call of $b_i=1$ at $c_i$ we can keep all of the outputs of this channel and bring them to $r_i$. 
This is because we are in the case of non-neighbouring calls, so the systems at $D_{i-1}$ and $D_{i+1}$ will not be returned to the referee, so it is never necessary to send systems from $c_i$ to $r_{i-1}$ or $r_{i+1}$.
On the other hand, when there is a call of $b_i=0$ to $D_i$, some systems will be sent to $D_{i+1}$ and some to $D_{i-1}$. 

Consider an isometric extension of the channel applied at each site, which can be conditioned on the bit given there. Call these isometries $V^{0}_{X_i\rightarrow Y_{i,-}Y_{i,+}}$ and $V^1_{X_i\rightarrow Z_i}$. 
$Y_{i,-}$ labels the system that will be sent to $D_{i-1}$ when a 0 call is received; $Y_{i,+}$ labels the system that will be sent to $D_{i+1}$ when a 0 call is received.
When a 1 call is received all the outputs can be sent to $r_i$, so we consider only a single output system $Z_i$. 
Consider the state
\begin{align}
    \ket{\Psi^0}_{Y} = \bigotimes_{i=1}^{5} V^0_{X_i\rightarrow Y_{i,-}Y_{i,+}}\ket{\Psi}_{X}
\end{align}
Rather than distribute $\ket{\Psi}$, we can equivalently distribute this state with $Y_{i,-}Y_{i,+}$ given to $D_i$. 
This is equivalent to giving $X_i$ from $\ket{\Psi}$ to $D_i$ since if a $0$ call is received at point $c_i$, the correct operation is already applied, while if a $1$ call is received at $c_i$ the operation $V^0_{X_i\rightarrow Y_{i,-}Y_{i,+}}$ can be inverted and then $V^1_{X_i\rightarrow Z_i}$ applied, which implements the original protocol. 

Now we can notice that $\ket{\Psi^0}_Y$ defines an entanglement sharing scheme.
In particular, correctness of the protocol requires that each set $T_i=Y_{i-1,+}Y_{i,-}Y_{i,+}Y_{i+1,-}$ is in an authorized pair with $T_j=Y_{j-1,+}Y_{j,-}Y_{j,+}Y_{j+1,-}$ for $j=i+2, i+3$.
Further, there is no data at $D_i$ that indicates if the partner will be $D_{i+2}$ or $D_{i+3}$, so this is an entanglement sharing scheme with an unknown partner. 
This allows us to apply \cref{lemma:nooddcyclesESS}, so that pair access structures with odd cycles are forbidden. 
For our needed access structure there is an odd cycle: $\{T_1,T_3\}$, $\{T_3,T_5\}$, $\{T_5,T_2\}$, $\{T_2,T_4\}$, $\{T_4,T_1\}$. 
Thus there can be no such protocol. 

\section{Discussion}

In this work we defined and studied the notion of an entanglement sharing scheme. 
We partially characterized for which access structures entanglement sharing schemes can be implemented, in both the cases of a known and unknown partner. 
Concretely, we gave a list of necessary conditions in both cases that we conjecture are also sufficient, and showed they are sufficient under a weakened notion of security. 
We also gave if and only if conditions for the natural, strong, notion of security in the case where we restrict to stabilizer encodings. 

We believe the entanglement sharing setting gives a natural framework within which to explore the patterns of entanglement that can be realized in quantum states. 
As well, ESS has already found one application (to entanglement summoning) and seems likely to find further applications. 
For instance, a distributed entanglement sharing state allows authorized pairs of parties to send and receive quantum messages (by preparing shared entanglement and then using teleportation), but does not allow unauthorized pairs to participate in the communication. 
A number of open questions present themselves in further developing the theory and applications of entanglement sharing. 

\vspace{0.2cm}
\noindent \textbf{Secret key vs. entanglement:} It would be interesting to understand a modified setting from the one we consider, where the goal is to have authorized pairs of parties able to distill secret, shared classical key, and unauthorized pairs to be unable to recover shared secret keys. 
For instance, in the $((2,2,5))$ construction pairs of type $\{2,2\}$ can distill key (for instance by first distilling shared EPR pairs) but pairings of type $\{1,2\}$ cannot. 
States that allow this flexible distillation of key may find interesting applications in network security. 

\vspace{0.2cm}
\noindent \textbf{Efficient schemes for efficient representations:} In the context of quantum secret sharing, some access structures beyond threshold schemes have efficient schemes, meaning the dimension of the system held by each party is not too large (for instance polynomial in the total number of parties). 
In particular, when whether a subset is authorized or unauthorized can be computed with a small span program, the scheme also has an efficient construction \cite{smith2000quantum}.
It would be interesting to explore if a similar statement is true for the bipartite settings discussed here. 

\vspace{0.2cm}
\noindent \textbf{Multipartite generalizations:} A generalization of our setting would be to consider the distribution of $k\geq 3$ party states within a scheme. 
For instance, we could have triplet access structures that consist of a set of authorized triplets and a set of unauthorized triplets. We could then ask for states that allow recovery of a fixed, entangled state from each authorized triple (e.g. recovery of a GHZ state) but the same state to not be well approximated by anything recovered from an unauthorized triple. 

\vspace{0.2cm}
\noindent \textbf{Entanglement sharing and entanglement summoning:} Inspired by the application of entanglement sharing schemes to the entanglement summoning problem given in \cref{fig:pentagonintro}, we can ask if ESS can be applied more generally to entanglement summoning as defined in \cite{adlam2018relativistic, dolev2021distributing}.\footnote{After the initial appearance of this work, this was taken up in \cite{bozanic2026entanglement}.}

\vspace{0.2cm}

\noindent \textbf{Acknowledgements:} We thank Caroline Sim\~{o}es who was involved in the early discussions which led to this project. 
AM, DL, and SM acknowledge the support of the Natural Sciences and Engineering Research Council of Canada (NSERC); this work was supported by NSERC Discovery grants (RGPIN-2025-03966 and RGPIN-2024-03823), NSERC-UKRI Alliance grant (ALLRP 597823-24), and NSERC Alliance Consortia Quantum grants (ALLRP 578455-22). FS is supported by the European Commission as a Marie Skłodowska-Curie Global Fellow.
Research at the Perimeter Institute is supported by the Government of Canada through the Department of Innovation, Science and Industry Canada and by the Province of Ontario through the Ministry of Colleges and Universities.
This work is supported by the Applied Quantum Computing Challenge Program at the National Research Council of Canada and JSPS KAKENHI Grant Number 23KJ1154, 24K17047.

\appendix

\section{Quantum Reed–Solomon codes}\label{sec:RScodes}

\subsection*{Code Construction}

We start with the following Vandermonde matrix $V$, where $x_0,\cdots,x_{n-1}\in \mathbb{F}_p$ are distinct elements, $p$ is prime and $p \geq n$:
\begin{eqnarray}
V = \begin{pmatrix}
1 & 1 & 1 & \dotsb & 1 \\
x_0 & x_1 & x_2 & \dotsb & x_{n-1} \\
x_0^2 & x_1^2 & x_2^2 & \dotsb & x_{n-1}^{2} \\
\vdots & \vdots & \vdots & \vdots & \vdots \\
x_0^{n-1} & x_{1}^{n-1} & x_{2}^{n-1} & \dotsb & x_{n-1}^{n-1}
\end{pmatrix}.
\end{eqnarray}
Denote $W=(V^{-1})^{T}$.
We use the first $r$ rows from the matrix $V$ to define $X$ type stabilizers, and the last $n-k-r$ rows from the matrix $W$ to define $Z$ type stabilizers: 
\begin{eqnarray}
&S^{(X)}_i \equiv \otimes_{j=0}^{n-1} (X_j)^{V_{ij}}, \ &\text{for } i \in \{0, \dotsb, r-1\},\\
&S^{(Z)}_i \equiv \otimes_{j=0}^{n-1} (Z_j)^{W_{ij}}, \ &\text{for } i \in \{r+k, \dotsb ,n-1\} \ .
\end{eqnarray}
They mutually commute due to $VW^T={I}$. 
The code consists of $n$ physical qudits and $k$ logical qudits.

As a CSS code, the submatrices of $V$ and $W$ chosen above serve as the parity check matrices of $C_2$ and $C_1$ respectively (see \cref{sec:stabilizerstuff}).
Therefore, $C_2^\perp$ is generated by the first $r$ rows of $V$, while $C_1$ is generated by the first $(r+k)$ rows of $V$.
As in \cref{eq:CSSstate}, an unnormalized basis of the code space is given by
\begin{equation}
    \sum_{c\in \mathbb{F}_p^r} \ket{c_0\mathbf{V}_0+\cdots+c_{r-1}\mathbf{V}_{r-1} + s_0\mathbf{V}_r +\cdots+ s_{k-1}\mathbf{V}_{r+k-1}},
\end{equation}
where $\mathbf{V}_i$ are rows of the matrix $V$.
This is exactly \cref{eq:RSwavefunction}, which we copy here for convenience:
\begin{align}
    \ket{\bar{s}}\propto \sum_{c\in \mathbb{F}_p^r}\ket{f_{c,s}(x_0)}\otimes \ket{f_{c,s}(x_1)}\cdots\otimes\ket{f_{c,s}(x_{n-1})},
\end{align}
where $s\in\mathbb{F}_p^k$ and 
\begin{align}
    f_{c,s}(x)=c_0+c_1x+\cdots+c_{r-1}x^{r-1}+s_0x^r+\cdots+s_{k-1}x^{r+k-1}.
\end{align}

\subsection*{Logical Operators}

A basis of stabilizer operators and logical operators can be directly obtained from rows of the matrices $V$ and $W$.
We can arrange them as follows:
\[
\begin{array}{rcl}
\underbrace{\mathbf{V}_0, \ldots, \mathbf{V}_{r-1}}_{X\text{-stabilizer}}
& \underbrace{\mathbf{V}_r, \ldots, \mathbf{V}_{r+k-1}}_{X\text{-logical}}
& \mathbf{V}_{r+k}, \ldots, \mathbf{V}_{n-1} \\[6pt]
\mathbf{W}_0, \ldots, \mathbf{W}_{r-1}
& \underbrace{\mathbf{W}_r, \ldots, \mathbf{W}_{r+k-1}}_{Z\text{-logical}}
& \underbrace{\mathbf{W}_{r+k}, \ldots, \mathbf{W}_{n-1}}_{Z\text{-stabilizer}}
\end{array}
\]
Note that the condition $VW^T={I}$ ensures that  
the stabilizer generators commute with one another and with the logical operators, and $V_i W_i = \omega^{-1} W_i V_i$ for $i=r,\cdots,r+k-1$, so we have a valid stabilizer code.  

Next we prove some facts about the code distance, which were presented in the main text in \cref{sec:stabilizerstuff}. 

\begin{lemma}\label{lem:RSZ}
For the quantum Reed–Solomon code, any subset of $\ell$ qudits (where $r \leq \ell \leq r+k$) supports exactly $\ell-r$ independent $Z$-type logical operators.
\end{lemma}
In particular, any subset of $r + k$ qudits supports all the $Z$-logical operators up to equivalence, 
any subset of $r+1$ qudits supports a nontrivial $Z$-type logical operator;
no subset of $r$ qudits supports a $Z$-type logical operator.

\vspace{0.2cm}
\begin{proof}\,
A $Z$-type Pauli operator $\bar{Z}$ supported on $\ell$ qudits labelled by $\{a_j|j \in \{0,\dotsb, \ell-1\} \}$ has the following form:
\begin{equation}
\bar{Z} = \bigotimes_{j=0}^{\ell-1} (Z_{a_j})^{q_j}, 
\end{equation}
where $q=(q_0,q_1,\cdots,q_{\ell-1})^T$ varies over $\mathbb{F}_p^{\ell}$.
%
%
An operator is a logical operator for a stabilizer code if and only if it commutes with all stabilizer generators (see \cite{gottesmanthesis1997}). 
Thus $\bar{Z}$ is a logical operator 
if and only if it commutes with all $X$-type stabilizer generators.
Picking an $r\times \ell$ submatrix of $V$, defined as $\tilde{V}_{ij} \equiv V_{ia_j}$ ($0\leq i\leq r-1$, $0\leq j \leq \ell-1$), then the condition for $\bar{Z}$ to be a logical operator is: 
\begin{eqnarray}
\tilde{V}q = 0.
\end{eqnarray}
For $\ell \geq r$, the first $r$ columns of $\tilde{V}$ form an invertible $r\times r$ Vandermonde matrix, hence $\rank\tilde{V}=r$, hence 
\begin{equation}
    \dim\ker\tilde{V}=\ell-r.
\end{equation}

The operator $\bar{Z}$ is a logical identity operator (that is, a stabilizer operator) if and only if it commutes with all $X$-type stabilizers and logical operators.
Picking an $(r+k)\times \ell$ submatrix of $V$, defined as $\tilde{V}'_{ij} \equiv V_{ia_j}$ ($0\leq i\leq r+k-1, 0\leq j \leq \ell-1$), then the condition for $\bar{Z}$ to be a stabilizer operator is: 
\begin{eqnarray}
\tilde{V}'q = 0.
\end{eqnarray}
For $\ell \leq r+k$, the matrix $\tilde{V}'$ consists of $\ell$ columns of an invertible $(r+k)\times(r+k)$ Vandermonde matrix, and therefore $\rank \tilde{V}' = \ell$, implying $\dim \ker \tilde{V}' = 0$.
The number of independent $Z$-type logical operators is therefore
\begin{equation}
        \dim\ker\tilde{V}-\dim\ker\tilde{V}'=\ell-r.
\end{equation}
\end{proof}

\begin{lemma}\label{lemma:RSLX}
For a quantum Reed–Solomon code, any subset of $\ell$ qudits ($n-r-k \leq \ell \leq n-r$) supports exactly $\ell+r+k-n$ independent $X$-type logical operators.
\end{lemma} 
In particular, 
any subset of $n-r$ qudits supports all the $X$-logical operators up to equivalence,
any subset of $n-r-k+1$ qudits supports a nontrivial $X$-type logical operator;
no subset of $n-r-k$ qudits supports an $X$-type logical operator. 

\vspace{0.2cm}
\begin{proof}\,
All $X$-type stabilizers and logical operators can be represented as linear combinations of $\mathbf{V}_0,\cdots, \mathbf{V}_{r+k-1}$:    
\begin{eqnarray}
\bar{X} \leftrightarrow \sum_{i=0}^{r+k-1} b_i V_i,~~ b_i \in \mathbb{F}_p.
\end{eqnarray}
Suppose $\bar{X}$ is supported on at most $\ell$ qudits, then it is trivial on the remaining $\ell'=n-\ell$ qudits, denoted as $\{a_j|j \in \{0,\dotsb, \ell'-1\} \}$. 
Picking an $(r+k)\times \ell'$ submatrix of $V$, defined as $\tilde{V}_{ij} \equiv V_{ia_j}$ ($0\leq i\leq r+k-1, 0\leq j \leq \ell'-1$), then the condition is: 
\begin{eqnarray}
b^T\tilde{V}= 0.
\end{eqnarray}
For $\ell\geq n-r-k$, we have $\rank\tilde{V}=\ell'$.
Therefore, on the size-$\ell$ subset, the number of independent stabilizers and logical operators equals
\begin{equation}
    r+k-\ell'=r+k-n+\ell.
\end{equation}

Similarly, all $X$-type stabilizers can be represented as linear combinations of $\mathbf{V}_0,\cdots, \mathbf{V}_{r-1}$:    
\begin{eqnarray}
\bar{X} \leftrightarrow \sum_{i=0}^{r-1} b_i V_i,~~ b_i \in \mathbb{F}_p.
\end{eqnarray}
The condition that $\bar{X}$ is supported on $\ell$ qudits can be similarly represented as
\begin{eqnarray}
b^T\tilde{V}'= 0,
\end{eqnarray}
where $\tilde{V}'$ is a $r\times \ell'$ submatrix of $V$.
For $\ell\leq n-r$, we have $\rank\tilde{V}'=r$.
Therefore, the only stabilizer operator that is supported on the $\ell$ qudits is the identity operator. 

Therefore, there are $(\ell+r+k-n)$ independent $X$-type logical operators on any subset of size $\ell$.
\end{proof}

Combining the above lemmas, we see that the quantum Reed–Solomon code has code distance:
\begin{equation}
    d=\min\{d_X,d_Z\},~~d_Z=r+1,~~d_X=n-r-k+1.
\end{equation}

One may notice that the proofs of the above lemmas are very similar, essentially relying on the same key point: certain submatrices of a Vandermonde matrix have full rank.
In fact, the two lemmas are ``dual'' to each other.
For CSS codes, the cleaning lemma \cite{yoshida2010framework} states that
\begin{equation}
g_X(A) + g_Z(A^c) = k,
\end{equation}
for any subset $A$, where $g_X$ and $g_Z$ denote the numbers of independent $X$-type and $Z$-type logical operators, respectively.
It follows immediately that the two lemmas are in fact equivalent.

\section{Quantum secret sharing: an alternative construction}\label{app-QSS}

In this section, we reprove the following well-known result about quantum secret sharing:

\vspace{0.2cm}
\noindent\textbf{Theorem}
    \emph{An access structure is valid if and only if it satisfies monotonicity and no-cloning.}
\vspace{0.2cm}

We will prove this theorem by providing an alternative construction of a stabilizer state satisfying the QSS property. 
Compared to the constructions in \cite{gottesman2000theory, smith2000quantum}, our construction is arguably more straightforward.

Given an access structure $\A$, we pick the minimal elements that are not a superset of any others, and call the collection $\A_{\min}$.
We will first construct logical operators $X_A$ and $Z_A$ for each $A\in\A_{\min}$ such that:
\begin{equation}\label{eq:local93}
    [X_A,X_{A'}]= [Z_A,Z_{A'}]= 0,~~\{X_A,Z_{A'}\}=0,~~\forall A,A'\in\A_{\min}.
\end{equation}
We will specify a rule to assign qubits to parties and a rule to define the logical operators. 
Via our construction, each party may contain multiple qubits, and the number of qubits each party has may differ.
We use the following procedure:
\begin{itemize}
    \item Initialization: for each $A\in\A_{\min}$, assign one qubit for each party in $A$. Define $X_A=\otimes X$ where the tensor product is over the qubits just added. Define $Z_A$ similarly. 
    \item Anticommutation-1: For all $A\in\A_{\min}$ and $A'\in\A_{\min}$ such that $A'\neq A$:
    \begin{itemize}
        \item Pick any party $s\in A\cap A'$ and add a qubit to party $s$.\footnote{Such a party exists by the no-cloning hypothesis.}
        \item Extend $X_A$ by (tensor multiplying it with) the Pauli $X$ of the newly added qubit.
        \item Extend $Z_{A'}$ by the Pauli $Z$ of the newly added qubit.
    \end{itemize}
    \item Anticommutation-2: For each $A\in\A_{\min}$ such that $|A|$ is even, we arbitrarily pick a party $s\in A$ and extend $X_A$ and $Z_A$ by the Pauli $X$ and Pauli $Z$ respectively.
\end{itemize}
After this procedure, each $X_A$ will be a tensor product of Pauli $X$ operators, and each $Z_A$ will be a tensor product of Pauli $Z$ operators. 
This ensures the first equation in \cref{eq:local93} is satisfied.
Moreover, the following two properties hold:
\begin{enumerate}
    \item For each $A\in\A_{\min}$ and $A'\in\A_{\min}$ such that $A'\neq A$, $\supp(X_A)\cap\supp(Z_{A'})$ contains exactly one qubit.
    \item For each $A\in\A_{\min}$, $\supp(X_A)\cap \supp(Z_{A})$ has $|A|$ or $|A|+1$ qubits in it, and is always an odd number.\footnote{This is because all the qubits in $\supp(X_A)\cap \supp(Z_{A})$ were added during the initialization phase, or in anti-commutation-2 phase. The anti-commutation-1 phase never adds an $X_A$ and $Z_A$ operator to the same qubit.}
\end{enumerate}
These properties ensure the second equation in \cref{eq:local93} also holds.

We then define a (in general mixed) state $\rho$ via the following stabilizers:
\begin{equation}
    X_AX_\ell,~~Z_AZ_\ell~~(\forall A\in\A_{\min}).
\end{equation}
Here, $\ell$ is a reference qubit carrying the logical information.
The above operators mutually commute due to  \cref{eq:local93}.
Moreover, any other stabilizer that is nontrivial on $\ell$ must be a product of (an odd number of) the above stabilizers.
As a consequence of our initialization, where we assign separate qubits to parties for each authorized set they are a part of, its support always contains at least one $A\in \A_{\min}$.

The QSS is obtained via the Choi–Jamiołkowski isomorphism by viewing the state $\rho$ as a map from $\ell$ to the support of all subsets $A$. 
Under this isomorphism, the QSS has logical $X$ and $Z$ operators supported on subsystem $S$ exactly when the state $\rho$ has stabilizers of the form $X_\ell \mathcal{O}_S$ and $Z_\ell \mathcal{O}'_S$ for some $\mathcal{O}_S, \mathcal{O}'_S$. 
Since we have that $X_\ell X_A$ and $Z_\ell Z_A$ are stabilizers of $\rho$, this ensures that the encoded qubit can be recovered from every subset $A\in \mathcal{A}_{\text{min}}$ (\cref{lemma:stabdistillcondition}), and hence every $A\in \mathcal{A}$, so that the QSS scheme is correct. 
For security, recall that we noted above that every stabilizer with non-trivial support on $\ell$ must contain at least one $A\in \mathcal{A}_{\text{min}}$ in its support, so no further subsystems can contain $X$ and $Z$ logical operators.

\bibliographystyle{unsrtnat}
\bibliography{biblio}

\end{document}